\documentclass[10pt]{article}

\usepackage[letterpaper,margin=1in]{geometry}
\usepackage[T1]{fontenc}
\usepackage{times}
\usepackage{amsmath}
\usepackage{amssymb}
\usepackage{booktabs}
\usepackage{graphicx}
\usepackage[hidelinks]{hyperref}
\usepackage[round]{natbib}
\usepackage{caption}
\graphicspath{{figures/}}

\renewcommand{\bibinfo}[2]{}

\newcommand{\msun}{\ensuremath{M_\odot}}
\newcommand{\pmc}{\ensuremath{p_{\mathrm{MC\text{-}DRW}}}}
\newcommand{\pnst}{\ensuremath{p_{\mathrm{NST}}}}
\newcommand{\nmc}{\ensuremath{n_{\mathrm{MC}}}}

\title{A Cross-Band (X-ray\,$\times$\,Optical) Periodicity Search for Supermassive
Black Hole Binaries: A Null Result and the First Completeness-Corrected Constraint}

\author{
  Karan Akbari\,\thanks{\href{https://orcid.org/0009-0005-0550-4018}{ORCID 0009-0005-0550-4018}; \texttt{karanakbari14@gmail.com}}\\
  St.\ Xavier's College, Mumbai, India
}
\date{}

\begin{document}
\maketitle

\begin{abstract}
\noindent
We present the first sample-level search for supermassive black hole
binaries (SMBHBs) requiring coherent quasi-periodicity at a common period in
the X-ray and optical bands, over 1194 Swift-BAT hard X-ray AGN (Stage~1) and
175 4XMM-DR14 AGN (Stage~2). No source is a co-periodic candidate.
Each light curve is modelled as a damped random walk (DRW) and searched with a
Lomb--Scargle periodogram and a look-elsewhere-corrected Monte-Carlo significance.
Because DRW red noise is largely independent between corona and disc, we require
both bands individually significant with periods coincident within 5\%,
and gate the survivors with the model-independent
null-signal-template test of \citet{robnik2024}. Over $P=100$--$900$\,d the
completeness-corrected 95\% upper limit on the co-periodic fraction is
amplitude-dependent: $\lesssim3\%$ for hard-X-ray fractional modulation
$\gtrsim0.3$, $\approx15\%$ (precision-limited) at $0.2$, and uninformative below $\sim0.15$ (the
$\epsilon=1$ floor is $0.25\%$). The sensitivity is set by the hard X-ray monitoring, not the optical photometry or the statistics, the opposite of the usual assumption. Daily MAXI and RXTE/ASM monitoring of the brightest AGN raises the
X-ray completeness 5--8-fold, and the search remains null. Integrated over the BAT black-hole mass
function, the expected all-amplitude co-periodic fraction is
$\sim3\times10^{-2}f_{\rm bin}\delta_{\rm mod}$ (modulo a factor $g\!<\!1$), with $f_{\rm bin}$ the sub-pc binary fraction, $\delta_{\rm mod}$ the modulating duty cycle, and $g$ the fraction reaching recoverable hard-X-ray amplitude, so a
null is expected. We deliver a validated cross-band framework and
the first completeness-corrected constraint on the co-periodic fraction, ready for the denser X-ray monitoring of Einstein Probe and eROSITA.
\end{abstract}

\section{Introduction}
\label{sec:intro}

Galaxy mergers should drive central supermassive black holes (SMBHs) to
sub-parsec separations, forming bound binaries \citep{begelman1980} that are the
loudest individual sources of the nanohertz gravitational-wave (GW) background
now detected by pulsar-timing arrays \citep{agazie2023,antoniadis2023,reardon2023,xu2023}. Identifying the
electromagnetic counterparts of this population (which AGN host the binaries
producing the GW background) is a central open problem. A binary with orbital
period $P_{\rm orb}$ can modulate the accretion luminosity through periodic
accretion-rate variation or relativistic Doppler boosting of an orbiting
mini-disc \citep{dorazio2015}, producing observable periodicity at $P_{\rm orb}$
or a few times $P_{\rm orb}$.

Photometric periodicity searches have nominated of order $10^2$ candidates
\citep{graham2015,charisi2016,liu2019,chen2024}. The central difficulty is that
the stochastic ``red noise'' of AGN accretion, well described over much of the
optical/UV by a damped random walk \citep[DRW;][]{kelly2009,macleod2010},
routinely mimics a coherent signal over the few cycles that a finite baseline
samples \citep{vaughan2016}. When candidates are re-examined with significance
tests that correctly model the red-noise null, essentially none survive:
\citet{robnik2024} reassessed the 33 Palomar Transient Factory candidates of
\citet{charisi2016} with a model-independent null-signal-template (NST) test and
retained none, and \citet{huijse2025} found zero robust periodic AGN among
$3.8\times10^5$ Gaia DR3 sources.

Multi-wavelength coherence is a complementary test that has not yet been
exploited at the sample level. The hard X-rays of an AGN trace the compact
corona, whereas the optical continuum traces the outer accretion disc. Their
stochastic variability is, to first order, independent (and indeed observed
X-ray/optical correlations in AGN are surprisingly weak; see
\S\ref{sec:caveats}). But a genuine binary modulates the global accretion
flow and should imprint the same period on both bands. The motivation is purity. Single-band searches already reach the full sample. The red-noise false-positive tail limits them, and requiring a coincident period in
two quasi-independent channels multiplies their single-band false-alarm probabilities. In our sample this
coincidence requirement suppresses the optical red-noise false-positive excess (a
$\sim$40$\times$ tail of spurious flags; \S\ref{sec:results}) by a further
$\sim$10$^3$ (the $3.7\%$ chance period-coincidence rate times the X-ray flag rate;
\S\ref{sec:methods}), a discrimination no single light curve can reach.
Single-band searches in individual channels remain active and null or
near-null, each returning at most a handful of unconfirmed candidates: in
hard/soft X-rays \citep[BASS\,XVIII, a null on 941 BAT
AGN]{liu2020}, the contemporaneous Swift-BAT periodicity search of \citet{serafinelli2020},
and the eROSITA candidate search of \citet{tubinarenas2025}, and in the mid-IR
\citep[WISE periodic AGN]{luo2025}. Multi-band studies exist only for
individual pre-selected sources, and no systematic cross-band sample search has
been published. This paper builds and validates the first. It delivers a
calibrated, model-independent cross-band pipeline, the first
completeness-corrected constraint on the co-periodic AGN fraction, and the finding
that the X-ray monitoring cadence, not the optical data or the statistics, sets
the sensitivity of such searches, which points to where future effort should
go. We analyse
two samples: Stage~1 (Swift-BAT), which carries the primary result, and Stage~2
(serendipitous 4XMM-DR14), which is included to show that the pipeline
runs unchanged on a pointed X-ray archive, not to add an independent
constraint.

\section{Sample and data}
\label{sec:data}

\subsection{Stage 1: Swift-BAT hard X-ray AGN}
The X-ray parent sample is the Swift \citep{gehrels2004} Burst Alert Telescope
\citep[BAT;][]{barthelmy2005} 157-month survey \citep{lien2025}, which provides crab-weighted monthly $14$--$195$\,keV light
curves ($\sim$157 epochs, 2004--2017) for all detected sources. AGN are selected
from the catalogue \texttt{TYPE} classification using an explicit allow-list
(Seyfert sub-types, QSO, LINER, generic and ``other'' AGN, and the
jet-dominated classes BZQ/BZB/BZG/FSRQ/Beamed AGN), which recovers 1253 AGN.
The jet-dominated subset (182 sources; the other 1071 are accretion-driven) is carried separately so that a
per-mechanism limit can be reported, since the DRW noise model is best motivated
for accretion-driven variability. We caution that for jet-dominated sources the
hard X-ray and optical emission share a common (synchrotron/inverse-Compton) jet
origin and are intrinsically correlated. The cross-band independence that
gives the method its statistical power therefore does not hold there, and claimed
blazar periodicities are in any case dominated by red-noise artefacts unrelated
to sub-pc accretion-flow modulation. The jet-dominated limit should be read as a
single-mechanism constraint, and the accretion-driven subset is the physically
appropriate cross-band sample for this test.

\subsection{Optical: ZTF and ASAS-SN}
Optical light curves are drawn from the Zwicky Transient Facility public data
release \citep[ZTF $g,r$;][]{bellm2019,masci2019} via the IRSA cone-search service, and
from ASAS-SN Sky Patrol \citep[$V,g$;][]{shappee2014,kochanek2017}. ASAS-SN's near-all-sky
coverage provides a time baseline overlapping both BAT (2004--2017) and ZTF
(2018--). (We use the single most significant optical band per source; an explicit
cross-era period-persistence test is an extension we do not pursue here.)
We retain bands with $\geq 20$ clean epochs after flag cuts and $5\sigma$
outlier clipping. Requiring at least one usable optical band alongside the X-ray
light curve yields the final Stage~1 sample of 1194 AGN (1022
accretion-driven, 172 jet-dominated; the optical-coverage cut removes 59 of the
1253, including 10 jet sources).
The 20-epoch requirement is an inclusion floor, not a sufficiency claim. Each
band's MC-DRW null is simulated on that band's own epoch sampling, so its window
function and aliasing enter the null and the flag rate stays calibrated even on
sparse light curves. Sparse bands instead pay the cost in completeness: recovery
is near zero at 20--75 epochs (\S\ref{sec:validation}), and the completeness
correction carries that cost into the limit.

\subsection{Stage 2: 4XMM-DR14 serendipitous AGN}
To probe a fainter, higher-redshift population we use the 4XMM-DR14 catalogue
\citep[][]{webb2020} of XMM-Newton \citep{jansen2001} serendipitous detections, building per-source
X-ray light curves from EPIC band-8 ($0.2$--$12$\,keV) fluxes grouped by
\texttt{SRCID}. We keep sources with $\geq 20$ detections and cross-match
(within $5''$) to spectroscopic AGN in the Million Quasars (Milliquas)
catalogue \citep{flesch2023}, giving 178 matches; 175 have usable optical
coverage, none of them jet-dominated. XMM fluxes (erg\,cm$^{-2}$\,s$^{-1}$, $\sim10^{-13}$) are converted to
a magnitude-like log scale, $-2.5\log_{10}(F/\tilde F)$, so the same DRW priors
apply as for the optical bands.

\subsection{Light-curve preparation}
\begin{figure}[t]
\centering
\includegraphics[width=\textwidth]{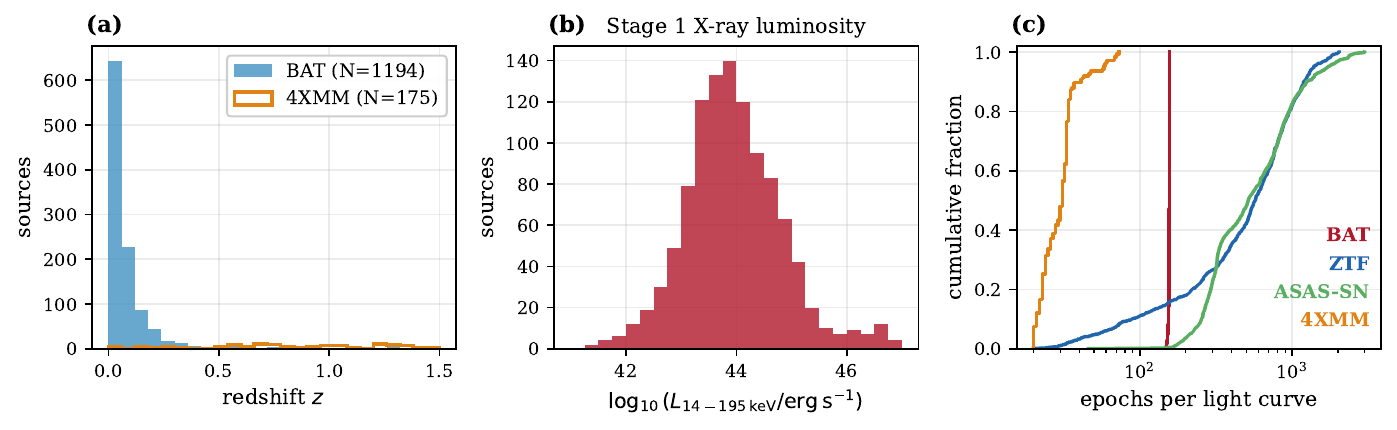}
\caption{Sample properties. (a) Redshift distribution of the Stage~1 BAT AGN and
the Stage~2 4XMM AGN; 4XMM reaches systematically higher $z$. Legend $N$ gives total sample sizes; the panel plots the 1085 BAT and 118 4XMM sources with catalogued redshift in the displayed range. (b) Hard X-ray
(14--195\,keV) luminosity of the Stage~1 sample. (c) Cumulative distribution of the number of epochs per light
curve, by survey (log $x$-axis): BAT is uniformly monthly, while the optical bands
span tens to thousands of epochs.}
\label{fig:sample}
\end{figure}

All light curves are stored as $(t,\,y,\,\sigma_y)$ with $y$ in magnitudes
(optical), count rate (BAT), or log-flux (XMM). Figure~\ref{fig:sample}
summarises the sample: redshift and hard X-ray luminosity distributions and the
per-band epoch counts. The BAT light curves are uniformly $\sim$monthly; the
optical bands range from tens to several thousand epochs.

\section{Methods}
\label{sec:methods}

For each (source, band) light curve we (i) estimate a DRW noise model, (ii)
compute a Lomb--Scargle (LS) periodogram on a masked log-frequency grid, (iii)
assign a look-elsewhere-corrected Monte-Carlo significance against a DRW null,
(iv) refine floor-saturated detections at high Monte-Carlo resolution, and then
combine the best X-ray and optical bands per source via Fisher's method and a
model-independent confirmation test.

\subsection{DRW noise model}
We adopt the MacLeod convention in which the DRW (Ornstein--Uhlenbeck) covariance
is $C(\Delta t)=\sigma^2\exp(-|\Delta t|/\tau)$, so the stationary variance is
$\sigma^2$ \citep[$\mathrm{SF}_\infty=\sqrt{2}\,\sigma$;][]{macleod2010}. Per
band we fit $(\log\tau,\log\sigma)$ by maximising the Gaussian-process marginal
likelihood (\texttt{celerite2}; \citealt{foremanmackey2017,foremanmackey2018}) with bounded,
multi-start L-BFGS-B, rejecting solutions that hit a parameter bound and falling
back to a method-of-moments (MoM) seed otherwise. Each fit's outcome is recorded
in a per-band \texttt{drw\_health} flag (a converged interior fit, a
bound-hitting fit, or a MoM fallback), which we use below to track the X-ray
DRW-fit pathology. Null realisations use the exact $O(N)$ Kelly recursion
\citep{kelly2009}.

\subsection{Periodogram and aliasing mask}
We compute the generalised LS periodogram \citep{lomb1976,scargle1982,vanderplas2018} on a log-spaced
grid of $n_f=750$ frequencies spanning observed-frame periods
$P\in[100,\,3000]$\,d, truncated so that the baseline contains at least $3$ complete
cycles (this suppresses the red-noise rise being read as a grid-edge period).
We mask the $1$-yr seasonal alias and its half-year harmonic
($355$--$380$\,d and $175$--$195$\,d for Stage~1; widened to
$340$--$390$\,d and $170$--$200$\,d for Stage~2; the $1$-yr alias is a known
systematic flagged by \citealt{charisi2016}). The
realised period sensitivity window is therefore $\approx 100$--$900$\,d.

\subsection{Monte-Carlo binned minimum-$p$ significance}
The periodogram is divided into 25 equal log-frequency bins. For the observed
light curve and for $\nmc$ DRW null realisations (same sampling, same
per-epoch errors, fitted $(\tau,\sigma)$ plus heteroscedastic measurement noise),
we record the maximum power in each bin. The per-bin survival function of the
null defines, for any realisation, a minimum-survival statistic
$T=\min_b s_b$; the global $p$-value is the fraction of null realisations with
$T$ at least as extreme as observed, floored at $1/(\nmc+1)$. This
look-elsewhere correction is at least as conservative as a per-bin Bonferroni
rule. LS power is correlated across neighbouring frequency bins, more so under
red noise, so the minimum-survival statistic is conservative; the sensitivity this
costs is absorbed into the injection-recovery completeness of \S\ref{sec:validation}
rather than biasing the upper limit. The broad pass uses $\nmc=2000$ (floor $\approx 5\times10^{-4}$).

\subsection{High-resolution refinement}
A source flagged in both bands at the broad pass may have hit the $\nmc=2000$
floor, so its $p$-values are only \emph{upper bounds}. Such sources are
re-measured at $\nmc=10^5$, which resolves the floor and frequently demotes a
band below the flag threshold (\S\ref{sec:refine}).

\subsection{Cross-band combination}
For each source we take the most significant X-ray band and the most significant
optical band. Because each source carries up to four optical bands (ZTF $g,r$;
ASAS-SN $V,g$), the selected optical $p$-value is the minimum over $k_{\rm opt}$
bands and is not Uniform under $H_0$. We therefore apply the look-elsewhere correction
$p_{\rm opt}\to1-(1-p_{\rm opt})^{k_{\rm opt}}$ before flagging and Fisher
combination. (The X-ray leg has a single band per source in the main BAT/XMM
pipeline, $k_X=1$, a no-op; the deep-cadence run of \S\ref{sec:deepcadence}
applies the analogous $k_X$ correction over \{BAT,\,MAXI,\,ASM\}.)
The joint significance is Fisher's combination \citep{fisher1925},
\begin{equation}
\begin{aligned}
X &= -2\left(\ln p_X + \ln p_{\rm opt}\right) \sim \chi^2_4 \quad (H_0),\\
p_{\rm joint} &= 1-F_{\chi^2_4}(X),
\end{aligned}
\end{equation}
which correctly accounts for combining two independent uniform $p$-values. The
naive product $p_X p_{\rm opt}$ is anti-conservative by a factor $|\ln p|$ at
small $p$ and is retained only as a diagnostic. A pair is ``period-matched'' if
$|P_X-P_{\rm opt}|/\bar P<0.05$; permuting the per-source X-ray and optical period
columns gives a chance coincidence rate of $\approx3.7\%$, so the period-match
requirement adds a $\sim$27-fold suppression on top of the two independent
single-band flags. The 5\% tolerance is a purity choice. Widening it to 10\%
raises the permutation chance rate to $7.2\%$ and the expected chance both-flagged,
period-matched count from $0.015$ to $0.029$, and it adds no completeness: all 73
both-band injection recoveries in the joint-completeness set (\S\ref{sec:jointcompl})
match within 5\%, none in the 5--10\% shell, and re-scoring the deep-cadence
recoveries at 5\% leaves every efficiency unchanged (\S\ref{sec:deepcadence}),
because a band that detects the injected signal recovers its period to better than
5\%. The 5\% cut costs no measurable completeness in our injection set and halves the chance-coincidence rate
relative to 10\%. Because a single very small $p$ can drive Fisher's
statistic, candidacy rests on the physical preconditions themselves: both bands
must be flagged at $\pmc<10^{-3}$ and the periods must match. We compute
$p_{\rm joint}$ only on sources that satisfy those preconditions, where it
summarises them and does not act as the gate.

\subsection{Model-independent confirmation (Robnik NST)}
Surviving candidates are confirmed with the null-signal-template test of
\citet{robnik2024}, implemented via the authors' \texttt{periodax} code. The
observed sinusoidal-template LS score is compared to the distribution of scores
under non-periodic templates whose per-cycle periods are randomised; we whiten
with the full DRW Cholesky factor so both templates see the correct coloured
null, and use $n_{\mathrm{NST}}=1000$. Because the NST is the final gate and
can only reject, never promote, it is applied only to bands whose source already
satisfies the both-flagged and period-match preconditions. A source that
fails those cannot become a candidate regardless of its NST $p$-value. (The full
broad NST pass over all flagged bands was run for Stage~1 and serves as the
methodology calibration; the NST is calibrated at $n_{\mathrm{NST}}=100$ and run at the
more conservative $n_{\mathrm{NST}}=1000$ in production. In the final cascade no source
reaches this gate, so the verdict does not depend on its exact false-alarm rate.)

\subsection{Candidate criteria}
A \emph{tier-1} candidate must satisfy, in order: (i) both the best X-ray and
best optical band flagged at $\pmc<10^{-3}$; (ii) period match within 5\%; and
(iii) both bands confirmed at $\pnst<10^{-2}$. \emph{Tier-2} relaxes the
confirmation gate to $\pnst<10^{-1}$. The Fisher $p_{\rm joint}$ is reported as a
summary of any survivors but is not itself a threshold: candidacy is defined by the
three physical preconditions, which keeps the criteria from depending on a number
($p_{\rm joint}$) that for two bands at the $10^{-3}$ gate already sits at
$\approx1.5\times10^{-5}$ and so is not independently reachable from that gate. The
operative single-band gate ($\pmc<10^{-3}$) lies just above the $\nmc=2000$
broad-pass floor ($5\times10^{-4}$), so a floor-saturated band passes it; the
measured broad-pass single-band false-positive rate at this gate is $1.0$--$1.3\%$
($\sim$10$\times$ nominal, from the calibration of \S\ref{sec:validation}: $1.05\%$
BAT, $1.00\%$ ZTF, $1.25\%$ ASAS-SN), so only candidate-path (both-flagged) bands
are re-resolved at $\nmc=10^5$, where the genuine significance is recovered.

\section{Validation}
\label{sec:validation}

\subsection{False-positive calibration} 
\begin{figure}[t]
\centering
\includegraphics[width=0.8\columnwidth]{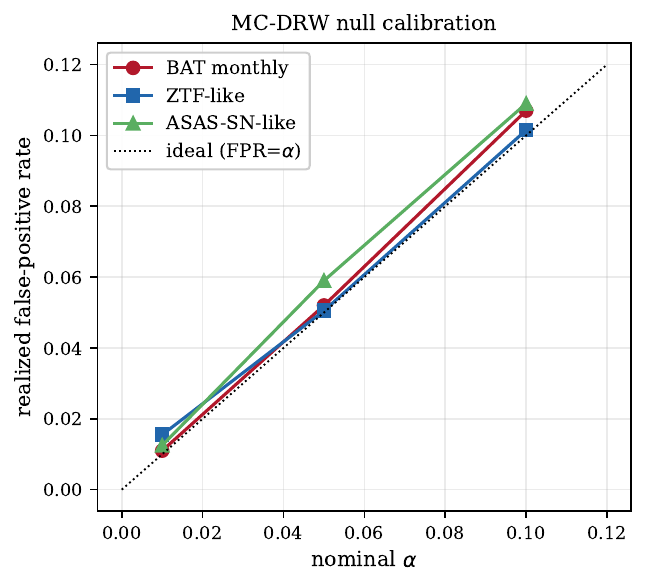}
\caption{False-positive calibration of the MC-DRW binned minimum-$p$ statistic on
pure-DRW simulations at the three survey cadences. The realised rate tracks the
nominal $\alpha$ (dotted ideal).}
\label{fig:fap}
\end{figure}

We calibrate the Monte-Carlo significance
on synthetic light curves at the three representative cadences. Drawing pure-DRW
realisations and recording $\pmc$, the realised false-positive rate tracks the
nominal $\alpha$ (Fig.~\ref{fig:fap}): $0.107,\,0.052,\,0.011$ at
$\alpha=0.1,\,0.05,\,0.01$ for the BAT cadence; the optical cadences are comparable
and mildly anti-conservative in the tail (realised $\approx0.012$--$0.016$ at
$\alpha=0.01$ for ASAS-SN/ZTF), which over-flags and so cannot manufacture a
false candidate.

\subsection{End-to-end completeness on real light curves} 
\begin{figure}[t]
\centering
\includegraphics[width=0.86\columnwidth]{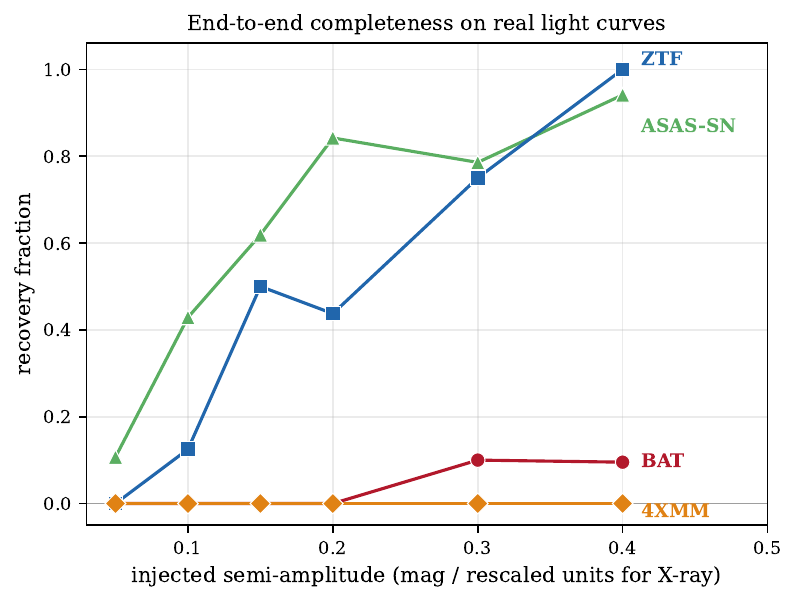}
\caption{End-to-end completeness from injecting signals onto real sample
light curves and running the identical fit$+$significance path. Optical bands
(ASAS-SN, ZTF) recover well above $\sim0.15$\,mag, whereas the hard X-ray BAT cadence
recovers only $\lesssim18\%$ and the sparse 4XMM cadence $\sim0\%$ (X-ray amplitudes
are in units of the rescaled, noise-dominated light-curve scatter). The $\lesssim18\%$ bound is the larger per-source BAT injection set; the plotted multi-band BAT curve peaks near $10\%$. This scatter-unit amplitude $a$ maps to a physical fractional amplitude $A=a\,s/\overline{\mathrm{rate}}$ (robust scatter $s$, mean count rate $\overline{\mathrm{rate}}$; \S\ref{sec:validation}).}
\label{fig:complreal}
\end{figure}

\begin{figure}[t]
\centering
\includegraphics[width=0.84\columnwidth]{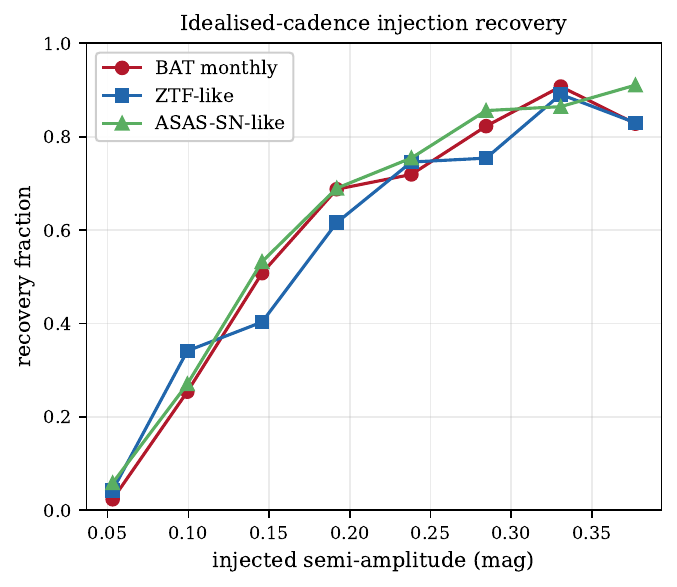}
\caption{Idealised-cadence injection-recovery completeness (synthetic data),
shown only for methodological comparison; the end-to-end real-light-curve
completeness (Fig.~\ref{fig:complreal}) supersedes it for the limit.}
\label{fig:completeness}
\end{figure}

A completeness curve
generated by feeding the true DRW parameters into the significance step
would bypass the per-band fit and therefore the dominant real-pipeline failure
mode. We instead inject known signals, a sinusoid and a relativistic
Doppler-boost mini-disc profile \citep{dorazio2015}, directly onto the
real sample light curves (960 injections) and run the identical
production path (per-realisation DRW re-fit $+$ MC-DRW), scoring recovery at the
same 5\% period tolerance candidacy requires (Fig.~\ref{fig:complreal}). The
result is strongly band-dependent: optical recovery (ASAS-SN, ZTF) rises to
$\approx50$--$90\%$ for semi-amplitudes $\gtrsim0.2$\,mag (rising to $\sim$1 at
$0.4$\,mag; Fig.~\ref{fig:complreal}), but the X-ray leg is the
binding constraint: the noise-dominated BAT monthly cadence recovers only
$\lesssim18\%$ (even at the largest amplitude) and the sparse 4XMM cadence
($\sim$20--75 epochs) essentially $0\%$. At a fiducial $\gtrsim0.2$\,mag-equivalent
amplitude the optical efficiency is $\epsilon_{\rm opt}\sim0.8$ and the X-ray
efficiency is the binding term. We quote two values of the latter, used
consistently throughout: $\epsilon_X\approx0.08$, the BAT-only recovery at
$\gtrsim0.2$\,mag measured on 471 well-sampled sources (the point value, pooled over $A=0.2$--$0.4$), and
$\epsilon_X\approx0.04$, the effective value once the faintest, most
sparsely-sampled BAT bands are folded in (the conservative value). The joint
per-source completeness is then $\epsilon_{\rm joint}=\epsilon_X\epsilon_{\rm opt}
\approx0.06$ (point) or $\approx0.03$ (conservative), a factorised estimate
superseded by the direct $\epsilon_{\rm joint}(A)$ below, which is the quantity
that enters the limit (\S\ref{sec:limits}). It is dominated by the hard X-ray
monitoring. A caveat on the factorisation and units: this $\epsilon_X\epsilon_{\rm opt}$
product assumes per-band recovery independence and (for the noise-dominated X-ray
bands) injects amplitude in rescaled-scatter units; we therefore also measure the
joint completeness directly, by co-injecting the same physical
fractional amplitude into a source's X-ray and optical bands and requiring both to
recover with matched period (reported below), which removes both
assumptions. Concretely, the X-ray axis of Fig.~\ref{fig:complreal} is the
injected sinusoid semi-amplitude $a$ in units of the rescaled BAT scatter: the
signal is added after the robust rescaling of \S\ref{sec:results}, so a value $a$
corresponds to a physical fractional amplitude $A=\delta F/\bar F=a\,(s/\overline{\mathrm{rate}})$,
with $s=\max(1.4826\,\mathrm{MAD},\,\mathrm{median}\,\sigma_i)$ the robust scatter
and $\overline{\mathrm{rate}}$ the mean count rate. For the noise-dominated BAT bands
$s\gtrsim\overline{\mathrm{rate}}$, so the two axes differ by that per-source ratio,
of order unity, and the direct co-injection of \S\ref{sec:jointcompl} sets $A$ in
physical units from the start. (For reference, the earlier idealised-cadence completeness of
Fig.~\ref{fig:completeness}, generated with synthetic data, is more optimistic
and is shown only for methodological comparison.) Restricting to BAT alone (i.e.\
the Stage-1 X-ray leg) the recovery is $\epsilon_X\approx0.08$ at $\gtrsim0.2$\,mag
over 1413 injections on 471 sources (a separate per-source X-ray injection set, distinct from the 960 multi-band injections above; pooled over $A=0.2$--$0.4$, rising $0.01\!\to\!0.17$ with amplitude); we further measure $\epsilon_X$ in bins of
X-ray luminosity and find it nearly flat ($\approx0.05$--$0.10$ across
$\log L_X\!\approx\!43$--$45$, varying by $<\!2\times$ with no monotonic trend in $L_X$), which we use in
\S\ref{sec:discussion} to justify the separability of the residence-completeness
estimate.

\subsection{Direct joint completeness and the amplitude-resolved limit}
\label{sec:jointcompl}
\begin{figure}[t]
\centering
\includegraphics[width=\columnwidth]{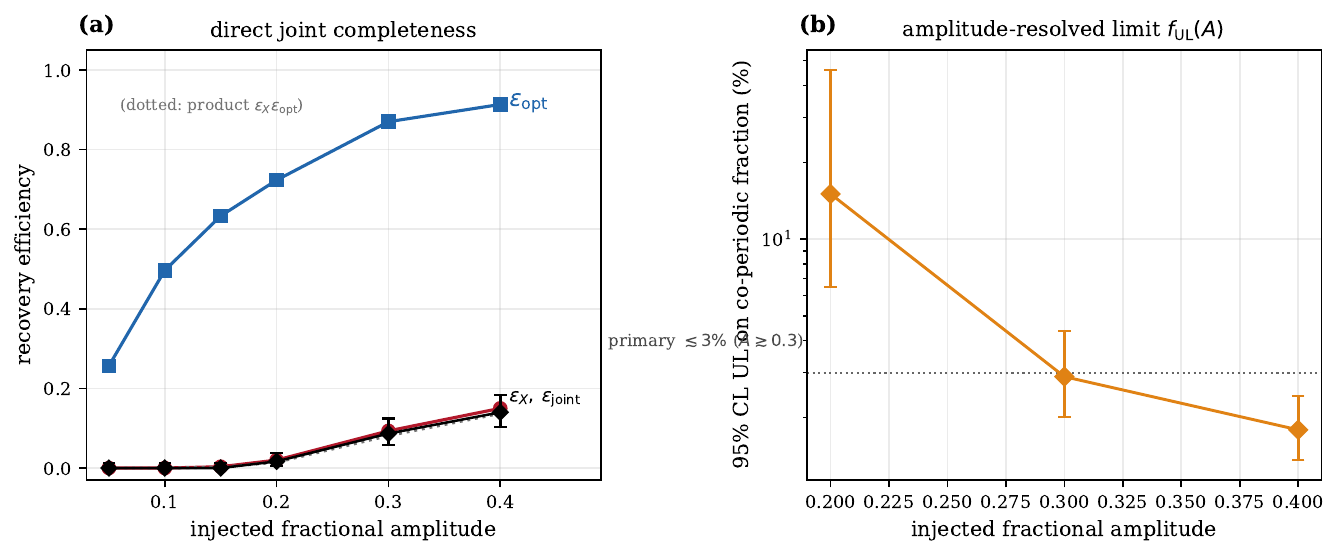}
\caption{Direct joint cross-band completeness from co-injecting the same physical
fractional amplitude into each source's X-ray and optical bands (1800 injections,
300 BAT$+$optical pairs). (a) Single-band and joint efficiencies vs amplitude; the
directly-measured $\epsilon_{\rm joint}$ tracks the product $\epsilon_X\epsilon_{\rm
opt}$ (dotted). (b) The resulting amplitude-resolved 95\% upper limit
$f_{\rm UL}(A)$: $\lesssim3\%$ for $A\gtrsim0.3$, weakening steeply below.}
\label{fig:joint}
\end{figure}

To remove both the factorisation ($\epsilon_X\epsilon_{\rm opt}$) and the
rescaled-unit ambiguity, we measure the joint completeness directly: into each of
300 sources' BAT and best-optical bands we co-inject the same sinusoid at a
common physical fractional amplitude $A\equiv\delta F/\bar F$ (optical $\delta m=2.5\,A/\ln10=1.086\,A$;
X-ray $\delta(\mathrm{rate})=A\,\overline{\mathrm{rate}}$). We run the production
pipeline on both and count a joint recovery only when both bands flag at
$\pmc<10^{-3}$ with peak periods within 5\% of the injection (1800 injections;
Fig.~\ref{fig:joint}; 300 BAT$+$best-optical pairs drawn at random, $\sim$300 per
amplitude bin, with injected periods spanning $\approx$140--800\,d, a subset of
the $100$--$900$\,d search window whose extreme edges are not probed). First, $\epsilon_{\rm joint}(A)$ is
steeply amplitude-dependent: $0.017$ at $A=0.2$ (Clopper--Pearson 95\% CI
$[0.005,0.038]$, $k=5$ recoveries), $0.09$ at $0.3$ ($k=26$), and $0.14$ at $0.4$
($k=42$), so a single completeness number is not meaningful and the limit must
be quoted as $f_{\rm UL}(A)$. Second, in physical fractional units the BAT recovery is
$\epsilon_X=0.02$--$0.15$ over $A=0.2$--$0.4$, confirming that the X-ray bottleneck is
physical and not an artefact of the rescaled-scatter axis. Third, the directly
measured $\epsilon_{\rm joint}$ agrees with the product $\epsilon_X\epsilon_{\rm
opt}$ within the binomial uncertainty (now consistent at all amplitudes, the
$A=0.2$ direct value resting on $k=5$ events), so the factorisation is a
reasonable approximation, although we use the direct measurement throughout. The
resulting limit is $f_{\rm UL}=\mu_{\rm UL}/(N\epsilon_{\rm joint}(A))$
(Fig.~\ref{fig:joint}b): a well-constrained $\lesssim3\%$ for fractional X-ray
modulation $A\gtrsim0.3$ (95\% CI $\sim2$--$4.4\%$), a precision-limited
$\approx15\%$ at $A=0.2$ (CI $\sim7$--$46\%$, $k=5$), and uninformative
below $A\sim0.15$ where
$\epsilon_{\rm joint}\to0$.

\subsection{Noise-model robustness} 
\begin{figure}[t]
\centering
\includegraphics[width=0.8\columnwidth]{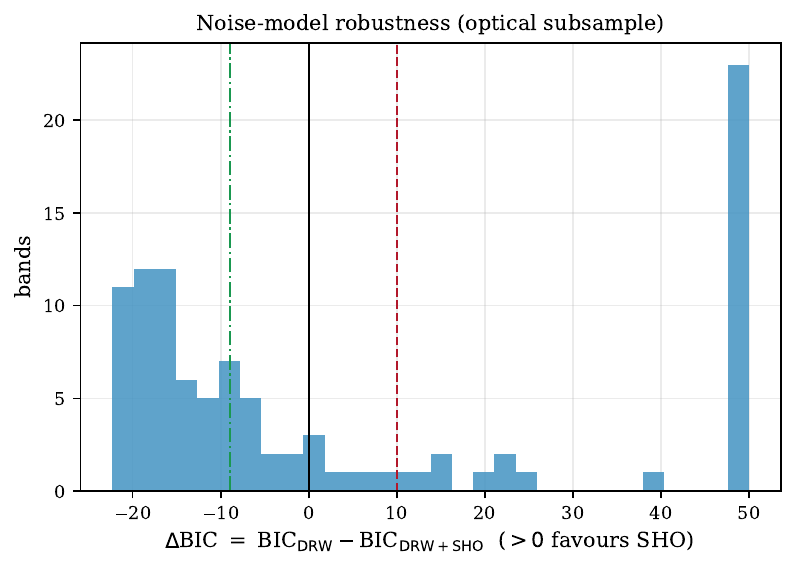}
\caption{Noise-model robustness: $\Delta$BIC between a plain DRW and a flexible
DRW$+$SHO kernel on 100 optical bands. Vertical lines mark $\Delta$BIC$=0$ (black), the median ($-9$, green dash-dot; favouring DRW), and the strong-preference threshold $\Delta$BIC$=+10$ (red dashed); the rightmost bin collects all $|\Delta$BIC$|\geq50$. The median favours DRW; the minority
preferring DRW$+$SHO carry excess high-frequency power, which only pushes the DRW
null further toward over-flagging.}
\label{fig:robust}
\end{figure}

Every $p$-value is computed against a DRW
null. To test sensitivity to PSD-shape mis-specification we re-fit a random
subsample of 100 optical bands with both a DRW and a more flexible DRW$+$SHO
(stochastically-driven harmonic-oscillator) kernel and compare by BIC
(Fig.~\ref{fig:robust}). The median $\Delta\mathrm{BIC}=-9$ favours the simpler
DRW, and DRW is adequate ($\Delta\mathrm{BIC}<6$) for $\sim67\%$ of bands; the
$\sim33\%$ that prefer DRW$+$SHO ($\sim31\%$ strongly, at $\Delta\mathrm{BIC}>10$)
carry extra high-frequency power and
would, if anything, make the DRW null over-flag, which is conservative
for the candidate selection and is the regime the model-independent NST gate
is designed to catch. This check is run on optical bands; the hard X-ray (BAT)
noise model, which carries the documented MoM-fallback pathology, is characterised
separately (\S\ref{sec:results}, \S\ref{sec:caveats}), where the same over-flagging
direction holds. The same DRW versus DRW$+$SHO comparison applied to the 1194
rescaled BAT light curves gives a median $\Delta\mathrm{BIC}=-13.5$: $99.6\%$ of
bands favour plain DRW and none strongly prefer the added SHO term
($\Delta\mathrm{BIC}>10$ for 0 of 1194; $\Delta\mathrm{BIC}>6$ for 3, the largest
$7.9$). The sparse monthly BAT sampling cannot resolve the extra high-frequency
term, so the plain-DRW null is adequate on essentially every BAT band. The null result does not depend on the DRW assumption.

\section{Results}
\label{sec:results}

\subsection{Single-band significance}
\begin{figure}[t]
\centering
\includegraphics[width=\columnwidth]{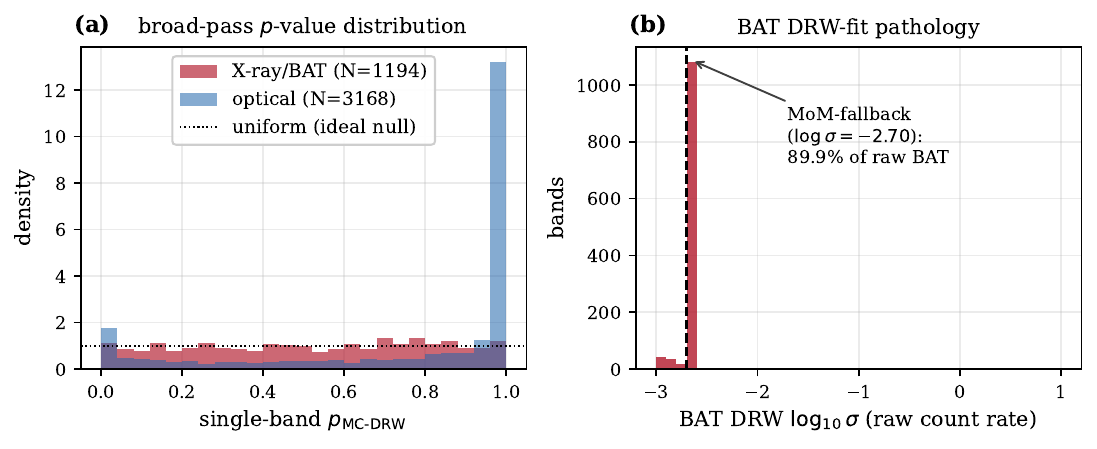}
\caption{(a) Broad-pass single-band $\pmc$ distributions for X-ray (BAT) and
optical bands in Stage~1; the optical distribution is approximately uniform apart
from a tail excess at $\pmc<10^{-3}$ ($\sim$40$\times$ the nominal rate, $\sim$4$\times$ the calibrated broad-pass rate; non-DRW
variability drives the excess above calibration), screened by the period-coincidence and NST
gates. (b) Distribution of the BAT DRW amplitude $\log\sigma$: 89.9\% pile up at
the method-of-moments fallback value $-2.70$, the signature of the count-rate
DRW-fit pathology that biases BAT toward over-flagging.}
\label{fig:pmc}
\end{figure}

Figure~\ref{fig:pmc}(a) shows the broad-pass single-band $\pmc$ distributions.
The optical distribution is approximately uniform apart from a tail excess at
$\pmc<10^{-3}$ ($\sim$40$\times$ the nominal $10^{-3}$ expectation; the calibrated broad-pass rate of \S\ref{sec:validation} accounts for a factor $\sim$10, non-DRW optical variability for the remaining $\sim$4), which is screened by the cross-band coincidence
and NST gates. The BAT distribution is more strongly skewed low, reflecting an
excess of nominal flags driven by the DRW-fit pathology of Fig.~\ref{fig:pmc}(b):
for the background-subtracted BAT count rates ($\sim10^{-3}$\,ct\,s$^{-1}$) the DRW
amplitude falls below the fit's $\sigma$ bound, and 89.9\% of BAT fits return the
clipped MoM seed ($\log\sigma=-2.70$) rather than a converged interior minimum.
This biases the BAT null toward over-flagging, with opposite consequences for the
two outputs of this paper (\S\ref{sec:caveats}). For the candidate verdict it is
conservative: an over-flagging X-ray leg makes the both-flagged$+$period-match
gate harder to pass cleanly, so a clean null is only strengthened. For the upper
limit it is optimistic: a wrong X-ray noise model degrades the true recovery
efficiency $\epsilon$ of a genuine co-periodic binary, and smaller $\epsilon$
makes the real limit weaker (larger) than quoted. We therefore re-fit all BAT
bands on a robust rescaling of the count rate, replacing each rate $r_i$ by
$(r_i-\tilde r)/s$ and its error by $\sigma_i/s$, where $\tilde r$ is the
light-curve median and $s=\max(1.4826\,\mathrm{MAD},\,\mathrm{median}\,\sigma_i)$
sets the scale to the robust scatter (floored at the typical error). This
restores converged interior fits for $\sim\!96\%$ of bands (the MoM-fallback fraction drops from $89.9\%$ to
$2.2\%$), and the broad-pass both-flagged count falls from $3$ to $1$, with the
last floor-saturated source resolved away by $\nmc=10^5$ refinement, leaving
$0$ both-flagged sources. LS periods are scale-invariant, so this changes no
period-match decision and the candidate count is unchanged. We fold the resulting (low,
X-ray-limited) completeness into the limit in \S\ref{sec:limits}.
The pre-rescale over-flagging is anti-conservative for the single-band flags,
and the $\nmc=10^5$ refinement and the NST gate screen exactly that: the
\S\ref{sec:refine} worked example is this mechanism in operation, a floor-saturated
MoM-fallback flag that dissolves once resolved. After the log-rate rescaling the
fallback fraction is $2.2\%$ ($26/1194$). Completeness is measured by injecting
through the identical fitting path, so the quoted $\epsilon_X$ carries the same
pathology as the flagging pipeline and stays self-consistent with it. A poorly fit
null costs sensitivity, which the completeness correction absorbs, and it does not
silently tighten the limit. Splitting the BAT injection recoveries by the source's
fit health gives a pooled recovery ($A\ge0.2$) of $0.082$ for healthy fits against
$0.033$ for MoM-fallback fits (10 fallback sources, single-digit recovery counts, so
directional rather than precise). Unhealthy fits under-recover. The injection set samples the same
health mix as the catalogue (10 of 471 sources, 2.1\%, against 2.2\% of the catalogue),
so the pooled $\epsilon_X$ already carries their deficit; assigning the healthy-fit
recovery to every source would raise $\epsilon_X$ by about 1\% relative, well inside its
binomial uncertainty. The pathology costs a small, measured amount of sensitivity and does
not bias the limit.

\subsection{Refinement removes the apparent candidate}
\label{sec:refine}
\begin{figure}[t]
\centering
\includegraphics[width=0.92\columnwidth]{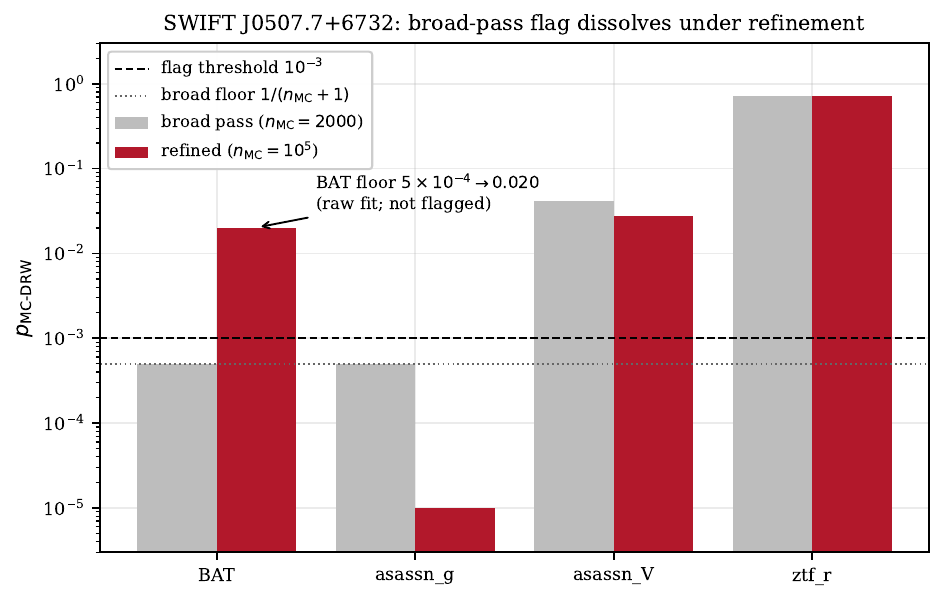}
\caption{The apparent candidate SWIFT\,J0507.7+6732. Broad-pass $p$-values
(grey) sit at the $\nmc=2000$ floor for the BAT and ASAS-SN\,$g$ bands; at
$\nmc=10^5$ (red) the BAT band resolves to $0.020$, above the $10^{-3}$ flag
threshold (dashed), removing the source from the candidate set. The refinement
step eliminates it before the model-independent test is reached.}
\label{fig:refine}
\end{figure}

At the broad pass one source, SWIFT\,J0507.7+6732 (the BL\,Lac 1ES\,0502+675,
$z=0.314$), appeared to satisfy the tier-1 preconditions: its BAT and ASAS-SN
$g$ bands both sat at the $\nmc=2000$ floor with a $1\%$ period match, giving an
apparent $p_{\rm joint}\sim4\times10^{-6}$. Refinement at $\nmc=10^5$
(Fig.~\ref{fig:refine}) resolves the BAT $p$-value from the floor
($5\times10^{-4}$) to $0.020$\footnote{This $0.020$ is the $\nmc=10^5$ refinement of the original raw count-rate BAT fit (the MoM-fallback regime; Fig.~\ref{fig:pmc}b). Under the robust count-rate rescaling adopted in \S\ref{sec:results} the same band gives $\pmc=0.29$, even further above the flag threshold, so the source is excluded under either treatment.} (a factor $\sim$20 above the flag threshold, $\sim$40 above the broad-pass floor),
so the source fails the both-flagged precondition and never reaches
confirmation. The apparent signal was a Monte-Carlo-resolution artefact on top
of an unreliable (MoM-fallback) BAT null. Broad-pass flags are upper bounds, and
significance is asserted only after refinement.

\subsection{Cross-band period coincidence: the null}
\begin{figure}[t]
\centering
\includegraphics[width=0.86\columnwidth]{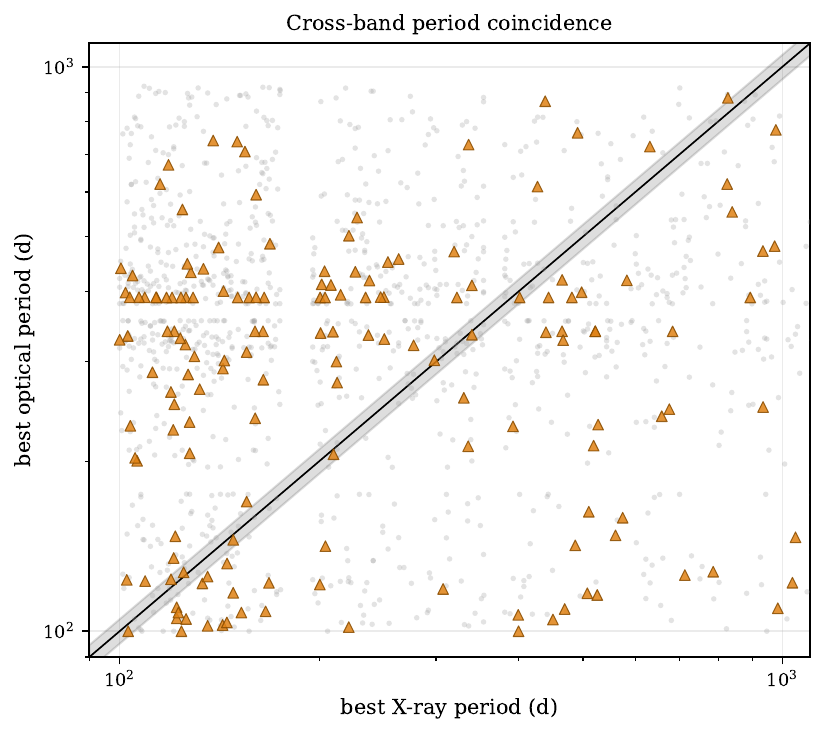}
\caption{Best X-ray period versus best optical period for all sources (Stage~1
grey circles, $N{=}1194$; Stage~2 4XMM orange triangles, $N{=}175$; solid line: equality). Open red symbols mark both-flagged
sources; the shaded band is the $|\Delta P|/\bar P<5\%$ period-match region. No
source lies in the both-flagged $\cap$ period-match region: there are no co-periodic
candidates.}
\label{fig:period}
\end{figure}

Figure~\ref{fig:period}, the central result, shows the best X-ray period against
the best optical period for every source, with the $\pm5\%$ period-match band
shaded. After the robust BAT re-fit and high-resolution refinement, neither
stage contains a single both-flagged source: the three sources that were both-flagged
under the raw count-rate BAT analysis were all driven by DRW-fit fallbacks (each
had a method-of-moments-fallback X-ray band, two also a fallback optical band,
flagged by the \texttt{drw\_health} diagnostic) and dropped out once the fits
converged; the one floor-saturated survivor (period mismatch $0.71$) was demoted
by $\nmc=10^5$ refinement. There are therefore 0 co-periodic sources, and
0 tier-1 and 0 tier-2 candidates, in either sample. The zero both-flagged,
period-matched outcome holds in each subsample on its own: 0 of the 1022
accretion-driven sources and 0 of the 172 jet-dominated sources, with the pipeline
run on the full 1194. Under the null the expected
number of chance both-flagged, period-matched sources across the sample is
$\approx0.015$ (14 X-ray and 34 LEE-corrected optical single-band flags among 1194 sources, times the $3.7\%$ chance period-coincidence rate), consistent with the zero observed: the cross-band
requirement suppresses the $\sim$40$\times$ optical red-noise excess to well below
one expected false candidate. The result is robust to the BAT treatment: the one both-flagged,
period-matched source that ever appeared (SWIFT\,J0507.7+6732, at the
original-sample broad pass; \S\ref{sec:refine}) was removed by $\nmc=10^5$
refinement, and the widened raw count-rate analysis contains no period-matched
both-flagged source at all. Table~\ref{tab:results} summarises the cascade.

\begin{table}[t]
\centering
\caption{Candidate cascade and 95\%-CL upper limits. Both-flagged counts are
after the robust BAT re-fit and $\nmc=10^5$ refinement. $f_{\rm UL}$ is the direct,
amplitude-resolved limit (Fig.~\ref{fig:joint}b); the $A=0.2$ value is
precision-limited ($k=5$ recoveries, 95\% CI $\sim7$--$46\%$). Stage~2 sets no
completeness-corrected fraction ($\epsilon_X\!\to\!0$ at its sparse cadence); its
$\epsilon{=}1$ rows are nominal floors only. Stage~2 contains no jet-dominated
sources, so its accretion-subset entry is the full Stage-2 sample.}
\label{tab:results}
\small
\begin{tabular}{lrr}
\toprule
& Stage 1 (BAT) & Stage 2 (4XMM)\\
\midrule
Sources (X-ray $+$ optical)      & 1194 & 175\\
Both-flagged ($\pmc<10^{-3}$)    & 0    & 0\\
Period-matched ($<5\%$)          & 0    & 0\\
NST-confirmed co-periodic        & 0    & 0\\
\textbf{Tier-1 candidates} ($\pnst<10^{-2}$) & \textbf{0} & \textbf{0}\\
Tier-2 candidates ($\pnst<10^{-1}$)          & 0          & 0\\
\midrule
UL, raw $\epsilon{=}1$ (optimistic floor)  & $0.25\%$ & $1.71\%$\\
$f_{\rm UL}$ at $A=0.2\,/\,0.3\,/\,0.4$ (direct) & $15\,/\,3\,/\,2\%$ & unconstr.\\
\textbf{UL, primary ($A\gtrsim0.3$)} & $\boldsymbol{\lesssim3\%}$ & unconstr.\\
\quad accretion subset ($\epsilon{=}1$)    & $0.29\%$ ($N{=}1022$) & $1.71\%$ ($N{=}175$)\\
\quad jet-dominated subset ($\epsilon{=}1$)& $1.74\%$ ($N{=}172$)  & $\cdots$\\
\bottomrule
\end{tabular}
\end{table}

\subsection{Stage 2}
With the log-flux conversion the XMM DRW fits converge normally, no source is
both-flagged, and the cross-band test returns 0 co-periodic sources. We caution
that the sparse 4XMM cadence ($\sim$20--75 epochs) makes this stage
nearly blind to the periods we target (\S\ref{sec:validation}): its periodicity
completeness is $\sim0$. This is compounded for the higher-redshift 4XMM sample by
time dilation: our search window is \emph{observed}-frame, so the rest-frame
orbital period is $P_{\rm obs}/(1+z)$, mapping to smaller separations and shorter
gravitational-wave residence, further suppressing the intrinsic Stage-2 yield
(negligible for the low-$z$ Stage-1 BAT sample). The clean Stage-2 null thus
shows that the framework ports without modification from an all-sky monitor (BAT)
to a serendipitous pointed archive (4XMM), which is the relevant test for applying
it to the better-sampled X-ray data sets (denser eROSITA or Einstein Probe
monitoring) that \S\ref{sec:discussion} identifies as the way forward; it does not
add a competitive constraint. The candidate-path
confirmation step had no bands to test in either stage: the outcome was settled
before the (expensive) model-independent stage was reached.

\subsection{Deep-cadence test: lifting the X-ray bottleneck for the brightest AGN}
\label{sec:deepcadence}

\begin{figure}[t]
\centering
\includegraphics[width=\textwidth]{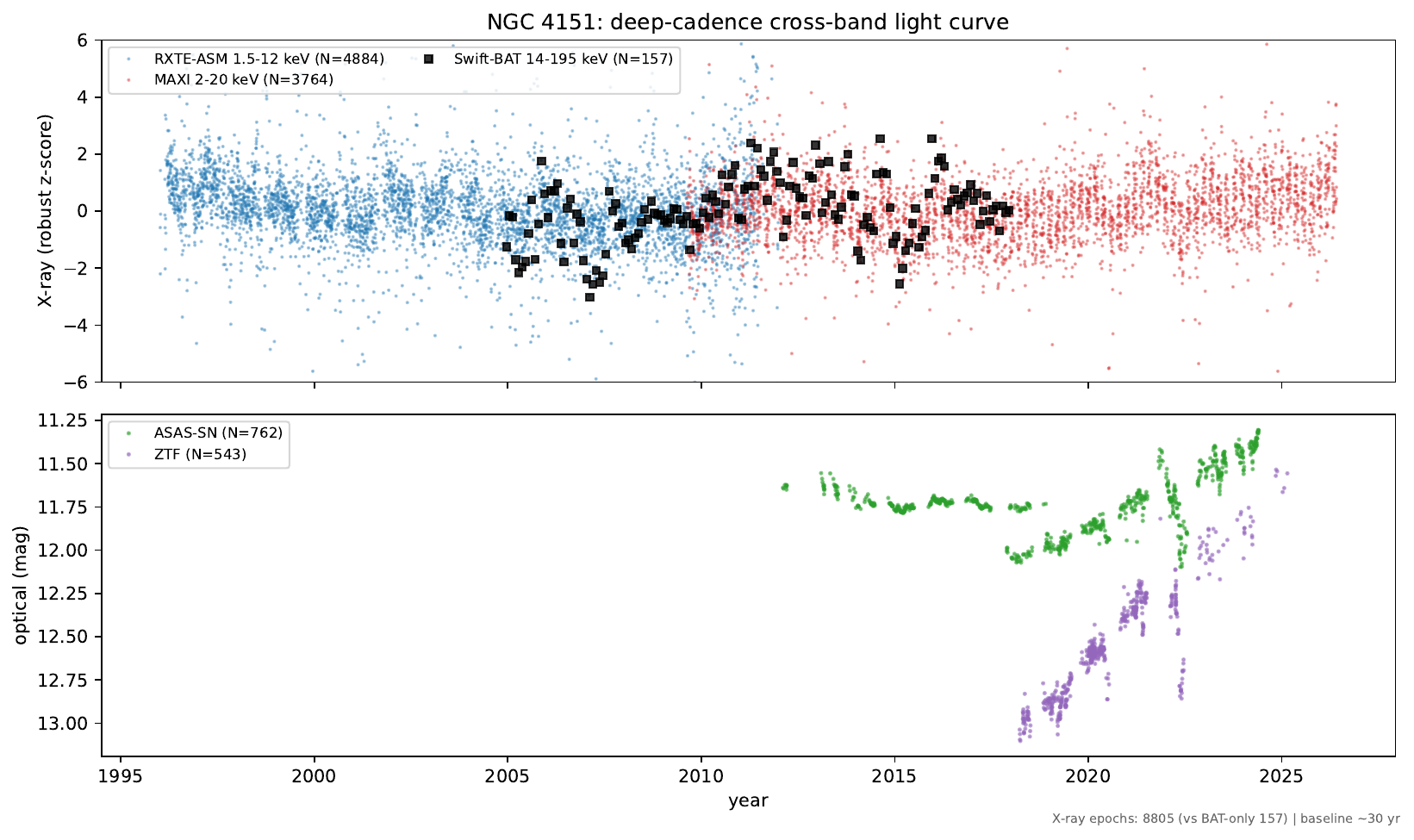}
\caption{The deep-cadence cross-band light curve of NGC\,4151. \emph{Top:}
the stitched hard/medium X-ray monitoring (RXTE/ASM, 1996--2012; MAXI,
2009--present; Swift-BAT) on a common robust scale, 8805 X-ray epochs over
$\sim$30\,yr versus 157 from BAT alone. \emph{Bottom:} the contemporaneous
ASAS-SN and ZTF optical photometry. For the X-ray-brightest AGN the monitors raise
the medium/hard X-ray completeness several-fold over BAT's
(\S\ref{sec:deepcadence}).}
\label{fig:deepcadence}
\end{figure}

The completeness analysis makes a testable prediction: where the hard X-ray
cadence is dense, the cross-band search should no longer be X-ray-limited.
We test this directly on the X-ray-brightest subsample,
for which archival all-sky monitors supply daily, decade-long light curves that
BAT's survey cadence cannot. For the 60 BAT AGN with the largest observed
$14$--$195$\,keV flux (all with existing optical coverage) we ingest two
additional X-ray bands: MAXI/GSC \citep{matsuoka2009} $2$--$20$\,keV daily light
curves (2009--present) and RXTE/ASM \citep{levine1996} $1.5$--$12$\,keV dwell
light curves (1996--2012), the latter inverse-variance binned to one point per
day. MAXI publishes a page for 40/60 of these sources and ASM a product
for 26/60, so 45/60 gain at least one monitor band (the misses are
chiefly Compton-thick Sy2s, absorbed out of the soft monitor band). Stitching ASM
and MAXI yields a near-continuous $\sim$30-yr daily X-ray baseline for the
brightest Seyferts: NGC\,4151, for example, reaches 8805 daily X-ray epochs
(4884 ASM $+$ 3764 MAXI $+$ 157 BAT) versus 157 from BAT alone, a $56\times$
increase, and samples $\sim$12--24 cycles inside the $100$--$900$\,d window
instead of BAT's noise-dominated single-cycle edge (Fig.~\ref{fig:deepcadence}).

The pipeline is run unchanged, with the monitors added to the X-ray survey set so
that the most significant of \{BAT, MAXI, ASM\} is taken per source. Because this
is a minimum over up to three (positively correlated) X-ray bands, we apply an
X-ray look-elsewhere correction $p_X\!\to\!1-(1-p_X)^{k}$, with $k$ the number of
X-ray bands searched for that source, before flagging and Fisher combination; this
is conservative (the bands are not independent). The optical leg here uses the raw
best-band $p$; the $k_{\rm opt}$ correction (\S\ref{sec:methods}) would only raise it,
so omitting it on this path is likewise conservative. The monitor count rates are
background-subtracted and frequently negative on faint sources, so they are put on
the same robust (median-absolute-deviation) scale as the BAT re-fit
(\S\ref{sec:results}) rather than a log scale. A residual 45\% of monitor bands
(35\% of MAXI, 62\% of the noisier ASM) still return the MoM-fallback DRW
amplitude, so their $p$-values rest on the same DRW null as BAT, with the same
over-flagging direction that is conservative for the candidate count. The ASM
daily errors are inflated to the empirical dwell-to-dwell scatter to forestall an
anti-conservative null on that monitor.

The result reproduces the global null in the regime where the search is
sensitive: across the 45 deep-cadence sources no source is both-flagged
and none is period-matched, giving 0 tier-1 and 0 tier-2 candidates. The
most significant monitor band in the whole sample (NGC\,1275, MAXI, $404$\,d,
$\pmc=0.012$) sits more than an order of magnitude above the $10^{-3}$ flag
threshold. The completeness changes substantially, though. We measure it with a
direct injection-recovery on the real monitor light curves: a sinusoid of physical
fractional amplitude $A\equiv\delta F/\bar F$ is added to the observed rate
(relative to the robust positive-flux scale), with an $A=0$ null control that
recovers nothing (false-recovery rate $0/24$). The pooled deep-cadence X-ray
efficiency is
$\epsilon_X=0.42,\,0.58,\,0.83$ at $A=0.2,\,0.3,\,0.4$ (Clopper--Pearson 95\% CIs
$[0.22,0.63]$, $[0.37,0.78]$, $[0.63,0.95]$ on 24 injections per amplitude over
6 MAXI $+$ 6 ASM monitor light curves at $\nmc=10^3$, scored at a looser 10\% period tolerance; MAXI $0.25/0.33/0.75$, the
longer-baseline ASM $0.58/0.83/0.92$). This is a factor of $\sim$5--8 above the BAT value
($\epsilon_X\!\approx\!0.02$--$0.15$) at $A\gtrsim0.3$, and more than $20\times$ at $A=0.2$.
These monitor efficiencies use a 10\% period tolerance and lower $\nmc$ than the
5\% main BAT measurement (\S\ref{sec:validation}); the lift exceeds what the
tolerance change alone can produce. Re-scoring the stored deep-cadence
recoveries at the main run's 5\% tolerance leaves the pooled efficiencies unchanged
at $0.42/0.58/0.83$: 46 of 47 detected injections recover within 5\%, and none fall
between 5\% and 10\%, so the disclosed tolerance mismatch does no work here.
Where the X-ray cadence is daily and decade-long, the
medium/hard X-ray leg therefore recovers a majority of injected modulations at
$A\gtrsim0.3$ ($\epsilon_X\!\approx\!0.4$--$0.8$) instead of a few per cent. The
cross-band coincidence requirement stops discarding real signals, and the null
over these sources is backed by measured sensitivity, unlike the latent null of
the full BAT sample. (The completeness lift is measured
directly on a 7-source injection subsample spanning the 12 monitor bands, 6 MAXI
and 6 ASM, and extrapolated across the 45, of which
19 have a monitor as their lowest-$p$ X-ray band.) But the resulting
fraction limit is not competitive: with only $N=45$
deep-cadence sources the factorised
$f_{\rm UL}=\mu_{\rm UL}/(N\,\epsilon_X\epsilon_{\rm opt})\approx27\%,\,19\%,\,13\%$
at $A=0.2,\,0.3,\,0.4$ ($\epsilon_{\rm opt}\!\approx\!0.6$, flagged as in
\S\ref{sec:jointcompl}) is weaker than the 1194-source Stage-1 limit
($\lesssim3\%$) because the well-monitored bright sample is small. The
value of the test is the demonstration that the cross-band method becomes
sensitive once the X-ray cadence is adequate; it does not provide an independent
population bound.

Three limitations apply to this test. (i) The monitored
sample is small and comprises the most-studied few dozen AGN in the sky, where an
obvious co-periodicity at these amplitudes would plausibly already be known; the
test is a confirmation in a high-completeness corner, not an independent blind
survey. (ii) It is restricted to the unobscured Seyferts: Compton-thick Sy2s
(NGC\,4945, the Circinus galaxy) are bright in BAT but absorbed out of the
$2$--$12$\,keV monitor band, and the jet-dominated blazars in the bright sample
(3C\,273, Mrk\,421, Mrk\,501) carry jet-driven variability and are reported in the
jet subset. (iii) The amplitude-ratio mechanism classifier is inoperative on the
rescaled monitor bands, and the ASM$+$MAXI baseline is a cross-instrument stitch
whose zero-point/scale seam could inject low-frequency power (a coherent period
across the seam would, conversely, be a strength). Within those limits, the same
search that is X-ray-completeness-limited over the full BAT sample becomes
X-ray-sensitive over the brightest few dozen AGN, and it remains null.

\section{Upper limits}
\label{sec:limits}

\begin{figure}[t]
\centering
\includegraphics[width=0.8\columnwidth]{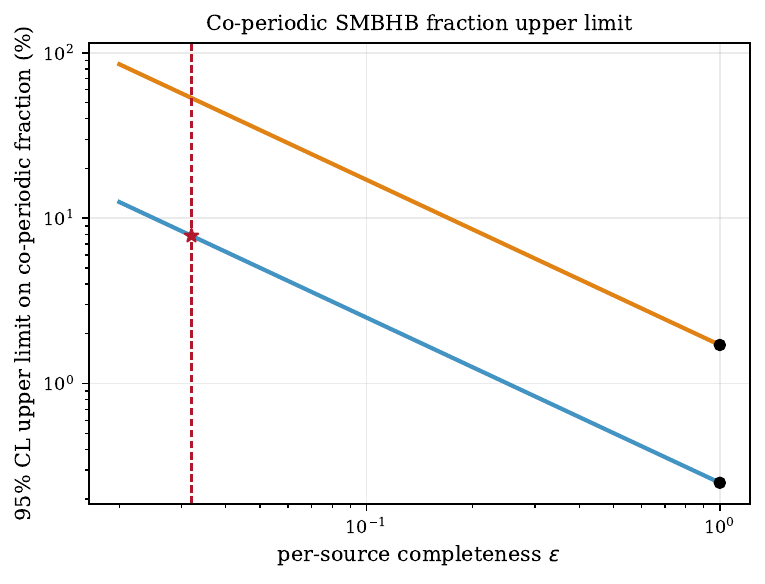}
\caption{Generic dependence of the 95\%-CL upper limit on the assumed per-source
completeness $\epsilon$ (illustrative). The upper (orange) curve is Stage~2 (4XMM, $N=175$) and the lower (blue) curve is Stage~1 (BAT, $N=1194$). Black points: the raw $\epsilon=1$ floor ($0.25\%$ and $1.71\%$ respectively);
red star: a factorised estimate $\epsilon_{\rm joint}\approx0.03$. The
amplitude-resolved $f_{\rm UL}(A)$ of
Fig.~\ref{fig:joint}(b) supersedes any single-$\epsilon$ reading.}
\label{fig:ul}
\end{figure}

With $n=0$ detections among $N$ searched sources at per-source completeness
$\epsilon$, the one-sided 95\%-CL Poisson upper limit on the mean number is
$\mu_{\rm UL}=-\ln(0.05)=2.996=\tfrac12\chi^2_{0.95}(2)\simeq3.0$ (the exact
value; \citealt{gehrels1986} is the conventional reference), so the
co-periodic-fraction limit is
$f_{\rm UL}=\mu_{\rm UL}/(N\epsilon)$. The choice of $\epsilon$ is decisive
(Fig.~\ref{fig:ul}). The raw $\epsilon=1$
values ($0.25\%$ for Stage~1, $N=1194$, and $1.71\%$ for Stage~2) are an
optimistic floor that assumes every co-periodic
binary in the sample would be recovered. The completeness-corrected number uses the directly-measured joint completeness
(\S\ref{sec:jointcompl}), which is steeply amplitude-dependent, so the limit is a
function of amplitude (Fig.~\ref{fig:joint}b),
$f_{\rm UL}(A)=\mu_{\rm UL}/(N\,\epsilon_{\rm joint}(A))$:
\begin{equation}
f_{\rm UL}\approx 15\%,\ 3\%,\ 2\%\quad\text{at}\quad A=0.2,\,0.3,\,0.4
\label{eq:ful}
\end{equation}
(fractional hard-X-ray modulation $A$; the $A\gtrsim0.3$ values are well-constrained,
95\% CI $\sim2$--$4.4\%$, while $A=0.2$ is precision-limited at the few-recovery level),
and uninformative below $A\sim0.15$ where $\epsilon_{\rm joint}\to0$. This
weakening is about sensitivity, not about the population: below $A\approx0.15$ the
surveys cannot detect the modulation even if every AGN hosted such a binary, so the
limit says nothing there. At $A\gtrsim0.3$ the constraint is data-limited rather
than completeness-limited. The
primary constraint is therefore $f_{\rm UL}\lesssim3\%$ for binaries with
hard-X-ray fractional modulation $\gtrsim0.3$. The weak low-amplitude tail reflects that the present hard X-ray
cadence cannot recover small modulations (averaging the direct
$\epsilon_{\rm joint}$ over $A\gtrsim0.2$ gives $\langle\epsilon_{\rm
joint}\rangle\approx0.08$ and $f_{\rm UL}\approx3\%$; a factorised
$\epsilon_X\epsilon_{\rm opt}$ cross-check, all in physical fractional units, is
consistent at $\sim3.4\%$). The limit carries two qualifications. (i) It is meaningful only where
$\epsilon_{\rm joint}$ is non-negligible, i.e.\ for large fractional X-ray
modulation. (ii) It applies to binaries with large amplitude in both
channels. The two amplitudes
need not match in general (disc/accretion-rate vs coronal modulation), but for the
relativistic Doppler-boost mechanism the band ratio can be calculated: the
fractional boost amplitude scales as $(3-\alpha_\nu)$ with spectral index
$\alpha_\nu$, so the hard X-ray ($\alpha_X\!\approx\!-1$, photon index
$\Gamma\!\approx\!2$) to optical ($\alpha_{\rm opt}\!\approx\!0$) ratio is
$(3-\alpha_X)/(3-\alpha_{\rm opt})\approx4/3$, so the X-ray amplitude is
modestly larger. Under boost, therefore, a binary detectable in optical is,
in amplitude, at least as detectable in hard X-ray, which reinforces that the
binding factor is X-ray cadence and noise, not an amplitude deficit. (We inject
equal fractional amplitude in both bands in the direct joint measurement,
which is conservative for the X-ray leg by this factor.) This conservatism is
mechanism-specific: for accretion-rate modulation the cross-band amplitude
ratio is unconstrained, so equal-amplitude injection could instead be optimistic,
and there the limit applies only to the subclass with comparable modulation in both
bands. The amplitude dependence
is dominated by the hard X-ray leg: optical completeness alone
would give $\sim0.3\%$ ($\epsilon_{\rm opt}\approx0.8$), so the steep $f_{\rm
UL}(A)$ is the X-ray recovery curve. We therefore report $f_{\rm UL}\lesssim3\%$
for $A\gtrsim0.3$ (Eq.~\ref{eq:ful}) as the Stage-1 constraint, with X-ray
monitoring as the binding limitation. Restricting to the
accretion-driven subset (the physically appropriate cross-band sample, $N=1022$)
with $\epsilon_{\rm joint}$ re-measured on the accretion-only injections gives an essentially identical limit, $f_{\rm UL}\approx15/3/2\%$
at $A=0.2/0.3/0.4$. For
Stage~2 the sparse 4XMM cadence gives $\epsilon_X\to0$, so no meaningful
completeness-corrected fraction can be set: the 4XMM stage is a methodology
demonstration, not a competitive constraint. The
expected number of chance co-periodic false positives across the sample is
$\approx0.015$ (\S\ref{sec:results}), confirming the search is not
threshold-limited.

\section{Caveats and limitations}
\label{sec:caveats}

\subsection{BAT DRW-fit pathology and its two-sided bias} As shown in
Fig.~\ref{fig:pmc}(b), 89.9\% of raw-count-rate BAT bands return the MoM-fallback
DRW amplitude because the background-subtracted rates lie below the fit's $\sigma$
bound. As quantified in \S\ref{sec:results}, this over-flagging bias has opposite
signs for the paper's two outputs (conservative for the candidate verdict,
optimistic for the $\epsilon=1$ limit), and the robust count-rate rescaling
adopted there restores converged fits while leaving the verdict at 0 candidates.
The $\epsilon=1$ limit (\S\ref{sec:limits}) should still be read as an optimistic
baseline for the X-ray leg until the end-to-end completeness (below) is folded in.

\subsection{DRW timescale bias} \citet{kozlowski2017} show that $\tau$ is
recovered without bias only when the baseline exceeds $\sim$10\,$\tau$; for the
$\sim$4700-d BAT baselines this means $\tau$ is trustworthy only below
$\sim$470\,d. Longer fitted $\tau$ should not be interpreted physically, and the
associated under-estimate of variance biases the null toward whiteness (over-flagging),
which only strengthens the null. Its effect on the limit is
already captured by the end-to-end completeness, which re-fits per realisation.

\subsection{Cross-band independence and its asymmetry} Fisher's combination
assumes the X-ray and optical $p$-values are independent. This is well motivated:
observed X-ray/UV--optical correlations in AGN are surprisingly weak (a known
tension for lamp-post reprocessing models), weaker still for the hard BAT band,
and the test combines periodicity $p$-values whereas any reprocessing
correlation is broadband. The strength of the argument is not
uniform across the two optical surveys, though. For BAT\,$\times$\,ZTF the epochs do not
overlap in time (BAT 2004--2017, ZTF 2018--), so instantaneous reprocessing
correlation is impossible. For BAT\,$\times$\,ASAS-SN, by contrast, the epochs
do overlap (ASAS-SN was included for that overlap;
\S\ref{sec:data}), so residual reprocessing correlation is in principle
possible for that pair, and the independence assumption is weakest there.
This is not a minority of the sample: ASAS-SN supplies the best optical band for
802/1194 sources (67\%; ZTF for 392, 33\%), so the overlapping-epoch case is the
majority. First,
the relevant correlation is in periodicity, not broadband flux, and a
genuine common period would be a signal, not a nuisance. Second, any positive nuisance
correlation only makes a both-band flag more likely under the null, so it is
conservative for our 0-candidate verdict, reinforced by the over-flagging X-ray
null above. Third, it matters only for sources that reach the precondition
stage, of which there are none with a period match. Quantitatively, even maximally correlated BAT and ASAS-SN nulls would raise the
both-flagged probability per source from $\sim10^{-6}$ to at most the single-band
gate $10^{-3}$, so $\lesssim1$ both-flagged source is expected across the sample,
and these must still period-match. With zero period-matched both-flagged sources
observed, the result is robust to any degree of cross-band correlation. A per-pair
correlated-noise treatment is a natural refinement for any future positive detection.

The two optical surveys also sample different eras from BAT. The ASAS-SN majority
overlaps the BAT baseline, but the 392 ZTF sources measure their optical period
mostly after the BAT window closes. A real SMBHB keeps the same period across
both eras, so a cross-era period match is still a valid test, weaker than a
contemporaneous one, and the permutation-calibrated chance rate already folds these
pairs in. The point bites only when interpreting a positive detection, where the
amplitude and phase can drift between eras. This search produced none, and an explicit
cross-era period-persistence test is the natural extension for a detection follow-up.

\subsection{Period range and completeness} The search is optical-baseline-limited
to $\approx100$--$900$\,d; classic long-period candidates (PG\,1302$-$102,
OJ\,287) lie outside it, and our null says nothing about them. The dominant
quantitative limitation is the X-ray completeness measured in
\S\ref{sec:validation}: the Stage-1 limit we quote is therefore the
completeness-corrected, amplitude-resolved $f_{\rm UL}(A)$ of
Eq.~\eqref{eq:ful} ($\lesssim3\%$ for fractional modulation $A\gtrsim0.3$) rather
than the $\epsilon=1$ floor of $0.25\%$. The limit
applies to co-periodic signals with a large fractional X-ray modulation
($A\gtrsim0.3$); below that the present hard X-ray cadence is uninformative.

\section{Discussion}
\label{sec:discussion}

Our null is consistent with the emerging picture that genuine periodic AGN
are rare: \citet{vaughan2016} argue most claimed periods are red-noise
artefacts, \citet{robnik2024} and \citet{huijse2025} retain none of the historical
optical candidates under model-independent or rigorous-null tests, and the
single-band BAT search of \citet{liu2020} (and the contemporaneous Swift-BAT periodicity
search of \citet{serafinelli2020}, with one $3.7\sigma$ candidate) is null, while
the eROSITA search of \citet{tubinarenas2025} reports 16 candidates flagged for
follow-up, not a null. But as we show below, our cross-band
sensitivity is currently X-ray-completeness-limited and so does not by itself
tighten that picture. The cross-band requirement is strictly stronger than any of
these single-band tests, so a cross-band null at a given amplitude sensitivity is
a correspondingly stronger statement about the same hypothesis class. To our
knowledge this is the first sample-level cross-band periodicity search: it
differs from the single-band BAT null of \citet{liu2020} by additionally
requiring optical co-periodicity at the X-ray period, and from the multiwavelength
analyses of individual binary candidates (e.g.\ PG\,1302$-$102; \citealt{dorazio2015}) by being a blind
population search, not a targeted follow-up of a pre-selected source. The work
delivers a validated cross-band framework, ready for
the denser X-ray monitoring that would activate its multiplied-false-alarm
advantage, and the empirical diagnosis that the X-ray completeness
currently limits such searches. Because the
X-ray leg dominates the (low) joint completeness, the astrophysical reach of
the present constraint is set by the quality of the available hard X-ray
monitoring, and the multiplied-false-alarm advantage of the cross-band
requirement becomes effective only once that monitoring improves.

\subsection{From a co-periodic fraction to a population statement} The
astrophysical content of a fractional limit follows from the separations probed,
and is steeply mass-dependent. At the observed periods $P=100$--$900$\,d
the orbital separation is only $\sim$1.4--6\,milliparsec (for total mass $M\sim10^{8.5}\,\msun$),
deep in the gravitational-wave-driven regime, where the circular coalescence time
\citep{peters1964}
$t_{\rm gw}=\tfrac{5}{256}\,c^5a^4/(G^3 m_1 m_2 M)\propto M^{-5/3}$ at fixed period
($a\propto M^{1/3}$). The window-crossing residence time
$\Delta t\approx t_{\rm gw}(900\,\mathrm{d})$ therefore swings by $\sim10^3$ across
the BAT mass range (equal-mass, from $\sim10^4$\,yr at $M_{\rm BH}\sim10^9\,\msun$ to
$\sim10^7$\,yr at $\sim10^7\,\msun$), so evaluating at a single mass is
misleading. Integrating the residence fraction $\Delta t/t_{\rm AGN}$ over the
BAT black-hole mass function \citep[the BASS DR2 active-BH mass function;][]{ananna2022},
which we approximate as log-normal with median $\log M_{\rm BH}\!\approx\!8.0$,
$\sigma\!\approx\!0.5$\,dex (individual masses from \citealt{koss2022}),
with a characteristic AGN lifetime $t_{\rm AGN}\sim10^8$\,yr \citep{marconi2004}
(identifying the measured AGN black-hole mass with the binary's \emph{total} mass) gives a sample-averaged residence fraction
$\langle\Delta t/t_{\rm AGN}\rangle\sim3\times10^{-2}$. This is a factor of several
($\sim$5$\times$) above the single-mass value at the MF median
($\log M\!\approx\!8.0$), because the long-$t_{\rm gw}$ low-mass tail dominates. The expected intrinsic co-periodic fraction (the fraction of AGN that
currently host a co-periodically modulating binary in our window, before any
detection efficiency) is the residence fraction weighted by the mass function,
\begin{equation}
f_{\rm exp}=f_{\rm bin}\,\delta_{\rm mod}\!\int\!\frac{\Delta t(M)}{t_{\rm AGN}}\,\phi(M)\,\mathrm{d}M
= f_{\rm bin}\,\delta_{\rm mod}\,\Big\langle\tfrac{\Delta t}{t_{\rm AGN}}\Big\rangle,
\label{eq:fexp}
\end{equation}
with $f_{\rm bin}$ the fraction of AGN ever hosting a sub-pc binary,
$\delta_{\rm mod}$ the modulating duty cycle, and $\phi(M)$ the mass function. The
integral gives $\langle\Delta t/t_{\rm AGN}\rangle\sim3\times10^{-2}$. Imposing a
gravitational-wave-validity cutoff that excludes the low-mass tail ($\log M_{\rm BH}\!\lesssim\!7$, where
$t_{\rm gw}(900\,{\rm d})\gtrsim3\times10^7$\,yr approaches $t_{\rm AGN}$) lowers this
by $\sim40\%$ to $\sim2\times10^{-2}$. The uncapped value is dominated by the
low-mass tail, where the GW-driven approximation is questionable, and is sensitive
to the assumed mass-function width and median; the capped $\sim2\times10^{-2}$ is
stable (it moves only over $0.018$--$0.025$ as the width is varied
$\sigma=0.4$--$0.6$\,dex, and over $0.013$--$0.034$ as the median is varied
$\log M_{\rm med}=8.2$--$7.8$), scales linearly with $t_{\rm AGN}$, and is the
more robust figure, order-of-magnitude either way. We compare
this intrinsic expectation to the completeness-corrected limit
(\S\ref{sec:limits}), which has the detection efficiency $\epsilon$ divided out,
so the two are directly comparable intrinsic fractions. That single-$\epsilon$
correction is valid across the mass-weighted population only if $\epsilon$ is
mass-independent. We verified that $\epsilon_X$ is nearly flat in $L_X$ (bin means
$0.05$--$0.10$, varying by $<\!2\times$ and non-monotonic in $L_X$; we bin by $L_X$ rather than $M$, lacking measured masses), so the
low-mass-dominated residence weighting and the detection efficiency do not strongly
anti-correlate. Two prefactors then matter:
$f_{\rm exp}\sim3\times10^{-2}\,f_{\rm bin}\delta_{\rm mod}$ is the
all-amplitude expected fraction, whereas our limit ($\lesssim3\%$) constrains
only the sub-population with recoverable hard-X-ray modulation $A\gtrsim0.3$, so the
like-for-like comparison multiplies $f_{\rm exp}$ by the (unknown) fraction $g$ of
co-periodic binaries reaching that amplitude. Only in the most optimistic corner, with
$f_{\rm bin}\delta_{\rm mod}\!\sim\!1$ and $g\!\sim\!1$ (nearly all co-periodic
binaries strongly modulating the hard X-ray), does the expected fraction
($\sim3\%$) become comparable to the limit, so the null marginally disfavours only
that corner. For any realistic $g\!<\!1$, or a plausible
$f_{\rm bin}\delta_{\rm mod}\!\sim\!10^{-1}$--$10^{-2}$, the expected fraction falls
$1$--$3$ orders below sensitivity and a null is firmly the expected outcome. Denser
X-ray monitoring (raising $\epsilon_X$, and thus $g$) closes the completeness
part of this gap but not the duty-cycle prefactor. The constraint is thus
a direct empirical bound consistent with nanohertz-background population models
\citep{agazie2023smbh,caseyclyde2022}. It does not yet test $f_{\rm bin}$ strongly.
Closing the gap needs a few-times-larger sample or, more effectively, the denser
X-ray monitoring the completeness analysis flags as the bottleneck. The two
SMBHB candidates singled out by the NANOGrav 15-yr targeted search (`Rohan' and
`Gondor'; \citealt{agarwal2026}) lie outside our X-ray archives and would require
dedicated pointings.

\subsection{The X-ray monitoring is the bottleneck} The end-to-end completeness
(\S\ref{sec:validation}) shows that the cross-band
method is presently limited by the X-ray cadence, not by the optical photometry or
the statistics:
optical recovery is $50$--$90\%$ at $\gtrsim0.2$\,mag, but the noise-dominated BAT
monthly cadence and the sparse 4XMM sampling recover periodicity in only a few per
cent of the joint injections at $A\sim0.2$, making the limit
amplitude-dependent ($\lesssim3\%$ for $A\gtrsim0.3$, weakening steeply below;
\S\ref{sec:limits}). The practical implication is that the gains from
a cross-band search will come from better-sampled X-ray light curves
(regular monitoring with, e.g., the Einstein Probe or continued eROSITA passes);
deeper optical data would add little. It also means the BAT$\times$optical and
4XMM$\times$optical combinations explored here are close to the floor of what hard
X-ray all-sky monitoring and serendipitous archives can deliver for periodicity.
We verify this directly in \S\ref{sec:deepcadence}: for the X-ray-brightest AGN,
substituting daily MAXI and RXTE-ASM monitoring for BAT's sparse cadence lifts the
medium/hard X-ray completeness several-fold, confirming that the bottleneck is
instrumental, not astrophysical, although the small ($N=45$)
well-monitored sample makes the resulting fraction limit a demonstration rather
than a competitive bound.

With $\epsilon_X\approx0.04$ the
X-ray leg adds little detection sensitivity at present cadence, so the cross-band
requirement currently costs completeness. What it still buys is purity: the
$\sim$10$^3$ false-alarm suppression quantified above (\S\ref{sec:results}) is why
the null is clean, not contaminated by red-noise candidates. The weak limit
reflects the present hard X-ray monitoring: we constrain $f_{\rm bin}\,\delta_{\rm mod}$ only weakly because the
experiment cannot yet see most of the signals it targets, and the deep-cadence
test (\S\ref{sec:deepcadence}) shows this completeness cost is instrumental and
lifts once the cadence improves. The cross-band advantage of multiplying
quasi-independent false-alarm rates becomes a tight constraint, not only a clean
null, once $\epsilon_X$ approaches the optical value.

The principal value of this work is the validated cross-band pipeline, together
with the demonstration (\S\ref{sec:refine}) that Monte-Carlo resolution, and not
only the model-independent test, eliminates spurious flags.
Future extensions (a longer optical baseline via CRTS to reach $P\sim2000$\,d, and
above all denser X-ray monitoring) are straightforward and do not alter the
present result.

\section{Conclusions}
\label{sec:conclusions}
\begin{enumerate}
\item We performed the first systematic, sample-level search for SMBHBs requiring
coherent X-ray$+$optical periodicity at a common period, at observed-frame periods
$P\approx100$--$900$\,d, over 1194 BAT AGN (Stage~1) and 175 4XMM AGN (Stage~2).
\item No source in either sample is a co-periodic candidate (0 tier-1,
0 tier-2), robust across accretion-driven and jet-dominated subsets (0 of the
1022 accretion-driven and 0 of the 172 jet-dominated sources).
\item A source flagged at low Monte-Carlo resolution was eliminated when refined,
demonstrating that broad-pass flags are upper bounds and must be resolved before
significance is claimed.
\item From a direct joint injection-recovery measurement (co-injecting the
same physical fractional amplitude into both bands), the completeness-corrected
95\%-CL Stage-1 limit on the co-periodic fraction is amplitude-dependent:
$\lesssim3\%$ for hard-X-ray fractional modulation $\gtrsim0.3$ ($\lesssim2\%$ at
$\gtrsim0.4$), weakening to $\approx15\%$ (precision-limited) at $0.2$ and
uninformative below $\sim0.15$. The raw $\epsilon=1$ value of $0.25\%$ ($N=1194$) is an optimistic floor,
not a conservative bound. The 4XMM stage sets no competitive fraction.
\item The search is X-ray-completeness-limited: the hard X-ray monitoring
is the binding constraint, ahead of the optical photometry or the statistics,
and the largest future gains will come from denser X-ray monitoring
(e.g.\ the Einstein Probe).
\item A direct test on the 60 X-ray-brightest AGN, for which daily MAXI and
RXTE-ASM monitoring replaces BAT's sparse cadence (a $\sim$56-fold gain in X-ray
epochs for NGC\,4151), confirms that the bottleneck is instrumental: the medium/hard
X-ray completeness rises several-fold and the cross-band search remains null. The
well-monitored sample is small ($N=45$), so this is a sensitivity-backed
demonstration that the method works where the X-ray cadence is adequate,
not a competitive fraction limit.
\item Integrating the (steeply mass-dependent, $t_{\rm gw}\propto M^{-5/3}$)
gravitational-wave residence time over the BAT black-hole mass function gives an
all-amplitude expected fraction $\sim3\times10^{-2}f_{\rm bin}\delta_{\rm mod}$ (GW-validity-restricted $\sim2\times10^{-2}$);
only in the most optimistic corner (high duty cycle and most binaries
modulating the hard X-ray at recoverable amplitude $A\gtrsim0.3$) is this comparable
to our $\lesssim3\%$ large-amplitude sensitivity, and it falls $1$--$3$ orders below
for realistic assumptions. Denser X-ray monitoring closes the completeness
part of the gap (raising $\epsilon_X$) but not the duty-cycle or amplitude
prefactors.
\end{enumerate}

\section*{Data and code availability}
\begin{sloppypar}
The analysis pipeline (DRW simulation/fitting in \texttt{periodogram.py} and
\texttt{drw.py}, the MC-DRW binned minimum-$p$ significance in
\texttt{significance/combined.py}, the Fisher cross-band combination in
\texttt{cross\_band.py}, the \texttt{periodax}-based NST wrapper in
\texttt{significance/robnik\_periodax.py}, the end-to-end driver
(\texttt{run\_v3\_redo.py}), the MAXI and RXTE-ASM ingest and deep-cadence driver
(\texttt{ingest/maxi.py}, \texttt{ingest/rxte\_asm.py}, \texttt{run\_bright\_agn.py},
\texttt{bright\_agn\_completeness.py}), and the figure
script \texttt{paper/make\_figures.py} that regenerates every figure herein from the result tables) and the result tables are archived on Zenodo, with a README documenting every file and column, at \href{https://doi.org/10.5281/zenodo.21386388}{doi:10.5281/zenodo.21386388}. All input catalogues are
public: Swift-BAT 157-month, ZTF (IRSA; \href{https://doi.org/10.26131/IRSA598}{doi:10.26131/IRSA598}, \citealt{irsa_ztf}), ASAS-SN Sky Patrol, 4XMM-DR14, Milliquas,
MAXI/GSC, and RXTE/ASM.
\end{sloppypar}

\section*{Acknowledgments}
I thank Sudip Bhattacharyya and Sayantan Bhattacharya for their guidance and for helpful discussions throughout this work.
Based on observations obtained with the Samuel Oschin Telescope 48-inch and the
60-inch Telescope at the Palomar Observatory as part of the Zwicky Transient Facility
project. ZTF is supported by the National Science Foundation under Grants No.\ AST-1440341
and AST-2034437 and a collaboration including current partners Caltech, IPAC, the Weizmann
Institute for Science, the Oskar Klein Center at Stockholm University, the University of
Maryland, Deutsches Elektronen-Synchrotron and Humboldt University, the TANGO Consortium of
Taiwan, the University of Wisconsin at Milwaukee, Trinity College Dublin, Lawrence Livermore
National Laboratories, IN2P3, University of Warwick, Ruhr University Bochum, Northwestern
University and former partners the University of Washington, Los Alamos National
Laboratories, and Lawrence Berkeley National Laboratories. Operations are conducted by COO,
IPAC, and UW.
This work made use of Swift-BAT, ZTF, ASAS-SN, XMM-Newton/4XMM, MAXI/GSC, RXTE/ASM,
and the Million Quasars catalogue.

This manuscript was prepared and edited with the assistance of Anthropic's
Claude \citep{claude2026}. All science, analysis, and conclusions are the author's own, and all
quantitative results are reproducible from the released pipeline scripts.

\noindent\textit{Facilities:} Swift (BAT), PO:1.2m (ZTF), ASAS-SN, XMM, MAXI, RXTE (ASM)

\noindent\textit{Software:} \texttt{astropy} \citep{astropy2013,astropy2018,astropy2022},
\texttt{numpy} \citep{harris2020}, \texttt{scipy} \citep{virtanen2020},
\texttt{matplotlib} \citep{hunter2007},
\texttt{celerite2} \citep{foremanmackey2017,foremanmackey2018},
\texttt{JAX} \citep{bradbury2018}, \texttt{astroquery} \citep{ginsburg2019},
\texttt{periodax} \citep{robnik2024}

\bibliographystyle{aasjournalv7}
\bibliography{refs}

\begin{thebibliography}{}
\expandafter\ifx\csname natexlab\endcsname\relax\def\natexlab#1{#1}\fi
\providecommand{\url}[1]{\href{#1}{#1}}
\providecommand{\dodoi}[1]{doi:~\href{http://doi.org/#1}{\nolinkurl{#1}}}
\providecommand{\doeprint}[1]{\href{http://ascl.net/#1}{\nolinkurl{http://ascl.net/#1}}}
\providecommand{\doarXiv}[1]{\href{https://arxiv.org/abs/#1}{\nolinkurl{https://arxiv.org/abs/#1}}}

% type= article
\bibitem[{N. {Agarwal} {et~al.}(2026){Agarwal} {et~al.}}]{agarwal2026}
{Agarwal}, N., {et~al.} 2026, \bibinfo{title}{{The NANOGrav 15 yr Data Set:
  Targeted Searches for Supermassive Black Hole Binaries},} \apjl, 998, L11

% type= article
\bibitem[{G. {Agazie} {et~al.}(2023{\natexlab{a}}){Agazie}
  {et~al.}}]{agazie2023}
{Agazie}, G., {et~al.} 2023{\natexlab{a}}, \bibinfo{title}{{The NANOGrav 15 yr
  Data Set: Evidence for a Gravitational-wave Background},} \apjl, 951, L8

% type= article
\bibitem[{G. {Agazie} {et~al.}(2023{\natexlab{b}}){Agazie}
  {et~al.}}]{agazie2023smbh}
{Agazie}, G., {et~al.} 2023{\natexlab{b}}, \bibinfo{title}{{The NANOGrav 15 yr
  Data Set: Constraints on Supermassive Black Hole Binaries from the
  Gravitational-wave Background},} \apjl, 952, L37

% type= article
\bibitem[{T.~T. {Ananna} {et~al.}(2022){Ananna} {et~al.}}]{ananna2022}
{Ananna}, T.~T., {et~al.} 2022, \bibinfo{title}{{BASS. XXX. Distribution
  Functions of DR2 Eddington Ratios, Black Hole Masses, and X-Ray
  Luminosities},} \apjs, 261, 9

% type= misc
\bibitem[{ {Anthropic}(2026){Anthropic}}]{claude2026}
{Anthropic}. 2026, {Claude (large language model)}, \url{https://claude.com};
  model versions Claude Opus~4 and Claude Fable~5 families; last accessed
  2026-07-16

% type= article
\bibitem[{J. {Antoniadis} {et~al.}(2023){Antoniadis} {et~al.}}]{antoniadis2023}
{Antoniadis}, J., {et~al.} 2023, \bibinfo{title}{{The second data release from
  the European Pulsar Timing Array. III. Search for gravitational wave
  signals},} \aap, 678, A50

% type= article
\bibitem[{ {Astropy Collaboration}(2013){Astropy Collaboration}}]{astropy2013}
{Astropy Collaboration}. 2013, \bibinfo{title}{{Astropy: A community Python
  package for astronomy},} \aap, 558, A33

% type= article
\bibitem[{ {Astropy Collaboration}(2018){Astropy Collaboration}}]{astropy2018}
{Astropy Collaboration}. 2018, \bibinfo{title}{{The Astropy Project: Building
  an Open-science Project and Status of the v2.0 Core Package},} \aj, 156, 123

% type= article
\bibitem[{ {Astropy Collaboration}(2022){Astropy Collaboration}}]{astropy2022}
{Astropy Collaboration}. 2022, \bibinfo{title}{{The Astropy Project: Sustaining
  and Growing a Community-oriented Open-source Project and the Latest Major
  Release (v5.0) of the Core Package},} \apj, 935, 167

% type= article
\bibitem[{S.~D. {Barthelmy} {et~al.}(2005){Barthelmy} {et~al.}}]{barthelmy2005}
{Barthelmy}, S.~D., {et~al.} 2005, \bibinfo{title}{{The Burst Alert Telescope
  (BAT) on the SWIFT Midex Mission},} Space Science Reviews, 120, 143

% type= article
\bibitem[{M.~C. {Begelman} {et~al.}(1980){Begelman}, {Blandford}, \&
  {Rees}}]{begelman1980}
{Begelman}, M.~C., {Blandford}, R.~D., \& {Rees}, M.~J. 1980,
  \bibinfo{title}{{Massive black hole binaries in active galactic nuclei},}
  \nat, 287, 307

% type= article
\bibitem[{E.~C. {Bellm} {et~al.}(2019){Bellm} {et~al.}}]{bellm2019}
{Bellm}, E.~C., {et~al.} 2019, \bibinfo{title}{{The Zwicky Transient Facility:
  System Overview, Performance, and First Results},} \pasp, 131, 018002

% type= misc
\bibitem[{J. {Bradbury} {et~al.}(2018){Bradbury} {et~al.}}]{bradbury2018}
{Bradbury}, J., {et~al.} 2018, {JAX: composable transformations of Python+NumPy
  programs}, \url{http://github.com/google/jax}

% type= article
\bibitem[{J.~A. {Casey-Clyde} {et~al.}(2022){Casey-Clyde}
  {et~al.}}]{caseyclyde2022}
{Casey-Clyde}, J.~A., {et~al.} 2022, \bibinfo{title}{{A Quasar-based
  Supermassive Black Hole Binary Population Model: Implications for the
  Gravitational Wave Background},} \apj, 924, 93

% type= article
\bibitem[{M. {Charisi} {et~al.}(2016){Charisi} {et~al.}}]{charisi2016}
{Charisi}, M., {et~al.} 2016, \bibinfo{title}{{A population of short-period
  variable quasars from PTF as supermassive black hole binary candidates},}
  \mnras, 463, 2145

% type= article
\bibitem[{Y.-C. {Chen} {et~al.}(2024){Chen} {et~al.}}]{chen2024}
{Chen}, Y.-C., {et~al.} 2024, \bibinfo{title}{{Searching for quasar candidates
  with periodic variations from the Zwicky Transient Facility: results and
  implications},} \mnras, 527, 12154

% type= article
\bibitem[{D.~J. {D'Orazio} {et~al.}(2015){D'Orazio}, {Haiman}, \&
  {Schiminovich}}]{dorazio2015}
{D'Orazio}, D.~J., {Haiman}, Z., \& {Schiminovich}, D. 2015,
  \bibinfo{title}{{Relativistic boost as the cause of periodicity in a massive
  black-hole binary candidate},} \nat, 525, 351

% type= book
\bibitem[{R.~A. {Fisher}(1925){Fisher}}]{fisher1925}
{Fisher}, R.~A. 1925, {Statistical Methods for Research Workers} (Edinburgh:
  Oliver and Boyd)

% type= article
\bibitem[{E.~W. {Flesch}(2023){Flesch}}]{flesch2023}
{Flesch}, E.~W. 2023, \bibinfo{title}{{The Million Quasars (Milliquas)
  Catalogue, v8},} The Open Journal of Astrophysics, 6, 49,
  \dodoi{10.21105/astro.2308.01505}

% type= article
\bibitem[{D. {Foreman-Mackey}(2018){Foreman-Mackey}}]{foremanmackey2018}
{Foreman-Mackey}, D. 2018, \bibinfo{title}{{Scalable Backpropagation for
  Gaussian Processes using celerite},} Research Notes of the American
  Astronomical Society, 2, 31

% type= article
\bibitem[{D. {Foreman-Mackey} {et~al.}(2017){Foreman-Mackey}
  {et~al.}}]{foremanmackey2017}
{Foreman-Mackey}, D., {et~al.} 2017, \bibinfo{title}{{Fast and Scalable
  Gaussian Process Modeling with Applications to Astronomical Time Series},}
  \aj, 154, 220

% type= article
\bibitem[{N. {Gehrels}(1986){Gehrels}}]{gehrels1986}
{Gehrels}, N. 1986, \bibinfo{title}{{Confidence Limits for Small Numbers of
  Events in Astrophysical Data},} \apj, 303, 336

% type= article
\bibitem[{N. {Gehrels} {et~al.}(2004){Gehrels} {et~al.}}]{gehrels2004}
{Gehrels}, N., {et~al.} 2004, \bibinfo{title}{{The Swift Gamma-Ray Burst
  Mission},} \apj, 611, 1005

% type= article
\bibitem[{A. {Ginsburg} {et~al.}(2019){Ginsburg} {et~al.}}]{ginsburg2019}
{Ginsburg}, A., {et~al.} 2019, \bibinfo{title}{{astroquery: An Astronomical
  Web-querying Package in Python},} \aj, 157, 98

% type= article
\bibitem[{M.~J. {Graham} {et~al.}(2015){Graham} {et~al.}}]{graham2015}
{Graham}, M.~J., {et~al.} 2015, \bibinfo{title}{{A systematic search for close
  supermassive black hole binaries in the Catalina Real-time Transient
  Survey},} \mnras, 453, 1562

% type= article
\bibitem[{C.~R. {Harris} {et~al.}(2020){Harris} {et~al.}}]{harris2020}
{Harris}, C.~R., {et~al.} 2020, \bibinfo{title}{{Array programming with
  NumPy},} \nat, 585, 357

% type= article
\bibitem[{P. {Huijse} {et~al.}(2025){Huijse} {et~al.}}]{huijse2025}
{Huijse}, P., {et~al.} 2025, \bibinfo{title}{{A search for periodic AGN
  variability in Gaia Data Release 3},} arXiv e-prints.
\newblock \doarXiv{2505.16884}

% type= article
\bibitem[{J.~D. {Hunter}(2007){Hunter}}]{hunter2007}
{Hunter}, J.~D. 2007, \bibinfo{title}{{Matplotlib: A 2D Graphics Environment},}
  Computing in Science and Engineering, 9, 90

% type= article
\bibitem[{F. {Jansen} {et~al.}(2001){Jansen} {et~al.}}]{jansen2001}
{Jansen}, F., {et~al.} 2001, \bibinfo{title}{{XMM-Newton observatory. I. The
  spacecraft and operations},} \aap, 365, L1

% type= article
\bibitem[{B.~C. {Kelly} {et~al.}(2009){Kelly}, {Bechtold}, \&
  {Siemiginowska}}]{kelly2009}
{Kelly}, B.~C., {Bechtold}, J., \& {Siemiginowska}, A. 2009,
  \bibinfo{title}{{Are the Variations in Quasar Optical Flux Driven by Thermal
  Fluctuations?},} \apj, 698, 895

% type= article
\bibitem[{C.~S. {Kochanek} {et~al.}(2017){Kochanek} {et~al.}}]{kochanek2017}
{Kochanek}, C.~S., {et~al.} 2017, \bibinfo{title}{{The All-Sky Automated Survey
  for Supernovae (ASAS-SN) Light Curve Server v1.0},} \pasp, 129, 104502

% type= article
\bibitem[{M.~J. {Koss} {et~al.}(2022){Koss} {et~al.}}]{koss2022}
{Koss}, M.~J., {et~al.} 2022, \bibinfo{title}{{BASS. XXII. The BASS DR2 AGN
  Catalog and Data},} \apjs, 261, 2

% type= article
\bibitem[{S. {Ko{\.z}lowski}(2017){Ko{\.z}lowski}}]{kozlowski2017}
{Ko{\.z}lowski}, S. 2017, \bibinfo{title}{{Limitations on the recovery of the
  true AGN variability parameters using damped random walk modeling},} \aap,
  597, A128

% type= article
\bibitem[{A.~M. {Levine} {et~al.}(1996){Levine}, {Bradt}, {Cui},
  {et~al.}}]{levine1996}
{Levine}, A.~M., {Bradt}, H., {Cui}, W., {et~al.} 1996, \bibinfo{title}{{First
  Results from the All-Sky Monitor on the Rossi X-Ray Timing Explorer},} \apjl,
  469, L33

% type= article
\bibitem[{A. {Lien} {et~al.}(2025){Lien} {et~al.}}]{lien2025}
{Lien}, A., {et~al.} 2025, \bibinfo{title}{{The 157 Month Swift/BAT All-sky
  Hard X-Ray Survey},} \apj, 989, 161

% type= article
\bibitem[{T. {Liu} {et~al.}(2019){Liu} {et~al.}}]{liu2019}
{Liu}, T., {et~al.} 2019, \bibinfo{title}{{Supermassive Black Hole Binary
  Candidates from the Pan-STARRS1 Medium Deep Survey},} \apj, 884, 36

% type= article
\bibitem[{T. {Liu} {et~al.}(2020){Liu} {et~al.}}]{liu2020}
{Liu}, T., {et~al.} 2020, \bibinfo{title}{{The BAT AGN Spectroscopic Survey.
  XVIII. Searching for Supermassive Black Hole Binaries in X-Rays},} \apj, 896,
  122

% type= article
\bibitem[{N.~R. {Lomb}(1976){Lomb}}]{lomb1976}
{Lomb}, N.~R. 1976, \bibinfo{title}{{Least-Squares Frequency Analysis of
  Unequally Spaced Data},} Astrophysics and Space Science, 39, 447

% type= article
\bibitem[{D. {Luo} {et~al.}(2025){Luo}, {Jiang}, \& {Liu}}]{luo2025}
{Luo}, D., {Jiang}, N., \& {Liu}, X. 2025, \bibinfo{title}{{A Systematic Search
  for Candidate Supermassive Black Hole Binaries Using Periodic Mid-infrared
  Light Curves of Active Galactic Nuclei},} \apj, 978, 86

% type= article
\bibitem[{C.~L. {MacLeod} {et~al.}(2010){MacLeod} {et~al.}}]{macleod2010}
{MacLeod}, C.~L., {et~al.} 2010, \bibinfo{title}{{Modeling the Time Variability
  of SDSS Stripe 82 Quasars as a Damped Random Walk},} \apj, 721, 1014

% type= article
\bibitem[{A. {Marconi} {et~al.}(2004){Marconi}, {Risaliti}, {Gilli}, {Hunt},
  {Maiolino}, \& {Salvati}}]{marconi2004}
{Marconi}, A., {Risaliti}, G., {Gilli}, R., {et~al.} 2004,
  \bibinfo{title}{{Local supermassive black holes, relics of active galactic
  nuclei and the X-ray background},} \mnras, 351, 169

% type= article
\bibitem[{F.~J. {Masci} {et~al.}(2019){Masci} {et~al.}}]{masci2019}
{Masci}, F.~J., {et~al.} 2019, \bibinfo{title}{{The Zwicky Transient Facility:
  Data Processing, Products, and Archive},} \pasp, 131, 018003

% type= article
\bibitem[{M. {Matsuoka} {et~al.}(2009){Matsuoka}, {Kawasaki}, {Ueno},
  {et~al.}}]{matsuoka2009}
{Matsuoka}, M., {Kawasaki}, K., {Ueno}, S., {et~al.} 2009, \bibinfo{title}{{The
  MAXI Mission on the ISS: Science and Instruments for Monitoring All-Sky X-Ray
  Images},} PASJ, 61, 999

% type= article
\bibitem[{P.~C. {Peters}(1964){Peters}}]{peters1964}
{Peters}, P.~C. 1964, \bibinfo{title}{{Gravitational Radiation and the Motion
  of Two Point Masses},} Physical Review, 136, B1224

% type= article
\bibitem[{D.~J. {Reardon} {et~al.}(2023){Reardon} {et~al.}}]{reardon2023}
{Reardon}, D.~J., {et~al.} 2023, \bibinfo{title}{{Search for an Isotropic
  Gravitational-wave Background with the Parkes Pulsar Timing Array},} \apjl,
  951, L6

% type= article
\bibitem[{J. {Robnik} {et~al.}(2024){Robnik} {et~al.}}]{robnik2024}
{Robnik}, J., {et~al.} 2024, \bibinfo{title}{{Periodicity significance testing
  with null-signal templates: reassessment of PTF's SMBH binary candidates},}
  \mnras, 534, 1609

% type= article
\bibitem[{J.~D. {Scargle}(1982){Scargle}}]{scargle1982}
{Scargle}, J.~D. 1982, \bibinfo{title}{{Studies in astronomical time series
  analysis. II. Statistical aspects of spectral analysis of unevenly spaced
  data},} \apj, 263, 835

% type= article
\bibitem[{R. {Serafinelli} {et~al.}(2020){Serafinelli}, {Severgnini}, {Braito},
  {et~al.}}]{serafinelli2020}
{Serafinelli}, R., {Severgnini}, P., {Braito}, V., {et~al.} 2020,
  \bibinfo{title}{{Unveiling Sub-parsec Supermassive Black Hole Binary
  Candidates in Active Galactic Nuclei},} \apj, 902, 10

% type= article
\bibitem[{B.~J. {Shappee} {et~al.}(2014){Shappee} {et~al.}}]{shappee2014}
{Shappee}, B.~J., {et~al.} 2014, \bibinfo{title}{{The Man behind the Curtain:
  X-Rays Drive the UV through NIR Variability in the 2013 Active Galactic
  Nucleus Outburst in NGC 2617},} \apj, 788, 48

% type= article
\bibitem[{D. {Tub{\'i}n-Arenas} {et~al.}(2025){Tub{\'i}n-Arenas}
  {et~al.}}]{tubinarenas2025}
{Tub{\'i}n-Arenas}, D., {et~al.} 2025, \bibinfo{title}{{Searching for
  supermassive black hole binaries within SRG/eROSITA-De. I. Properties of the
  X-ray selected candidates},} \aap, 698, A192.
\newblock \doarXiv{2505.02708}

% type= article
\bibitem[{J.~T. {VanderPlas}(2018){VanderPlas}}]{vanderplas2018}
{VanderPlas}, J.~T. 2018, \bibinfo{title}{{Understanding the Lomb-Scargle
  Periodogram},} \apjs, 236, 16

% type= article
\bibitem[{S. {Vaughan} {et~al.}(2016){Vaughan} {et~al.}}]{vaughan2016}
{Vaughan}, S., {et~al.} 2016, \bibinfo{title}{{False periodicities in quasar
  time-domain surveys},} \mnras, 461, 3145

% type= article
\bibitem[{P. {Virtanen} {et~al.}(2020){Virtanen} {et~al.}}]{virtanen2020}
{Virtanen}, P., {et~al.} 2020, \bibinfo{title}{{SciPy 1.0: fundamental
  algorithms for scientific computing in Python},} Nature Methods, 17, 261

% type= article
\bibitem[{N.~A. {Webb} {et~al.}(2020){Webb} {et~al.}}]{webb2020}
{Webb}, N.~A., {et~al.} 2020, \bibinfo{title}{{The XMM-Newton serendipitous
  survey. IX. The fourth XMM-Newton serendipitous source catalogue},} \aap,
  641, A136

% type= article
\bibitem[{H. {Xu} {et~al.}(2023){Xu} {et~al.}}]{xu2023}
{Xu}, H., {et~al.} 2023, \bibinfo{title}{{Searching for the Nano-Hertz
  Stochastic Gravitational Wave Background with the Chinese Pulsar Timing Array
  Data Release I},} Research in Astronomy and Astrophysics, 23, 075024

% type= misc
\bibitem[{ {ZTF Team}(2025){ZTF Team}}]{irsa_ztf}
{ZTF Team}. 2025, {ZTF Lightcurves}, IPAC,
  \url{https://doi.org/10.26131/IRSA598}; doi:10.26131/IRSA598

\end{thebibliography}

\end{document}